\documentclass[12pt]{iopart}

\usepackage{epsfig}
\usepackage[dvips]{color}

\def\alt{\raise0.3ex\hbox{$\;<$\kern-0.75em\raise-1.1ex\hbox{$\sim\;$}}}
\def\agt{\raise0.3ex\hbox{$\;>$\kern-0.75em\raise-1.1ex\hbox{$\sim\;$}}}
\def\d{{\rm d}}
\newcommand{\be}{\begin{equation}}
\newcommand{\ee}{\end{equation}}
\newcommand{\bea}{\begin{eqnarray}}
\newcommand{\eea}{\end{eqnarray}}

\definecolor{Black}{named}{Black}
\definecolor{Red}{named}{Red}

\newcommand{\bw}{\begin{widetext}}
\newcommand{\ew}{\end{widetext}}
\begin{document}
\def\d{{\rm d}}

\hfill DSF-36/2006, FERMILAB-PUB-06-478-A,  UCI-TR-2006-24

\title[The Signature of LSS on the VHE Gamma-Ray Sky]{The Signature of Large Scale Structures on the
Very High Energy Gamma-Ray Sky}

\author{A.~Cuoco$^{1}$, S.~Hannestad$^{2}$, T.~Haugb{\o}lle$^{2}$,
G.~Miele$^{1}$, P.~D.~Serpico$^{3}$, H.~Tu$^{2,4}$}

\address{$^1$ Dipartimento di Scienze Fisiche, Universit\`{a} di Napoli
$Federico$ $II$ and INFN Sezione di Napoli, Complesso
Universitario di Monte S.\ Angelo, Via Cinthia, I-80126 Napoli,
Italy.}

\address{$^2$ Institut for Fysik og Astronomi,
Aarhus Universitet Ny Munkegade, Bygn. 1520 8000 Aarhus Denmark.}

\address{$^3$ Center for Particle Astrophysics, Fermi National Accelerator
Laboratory, Batavia, IL 60510-0500 USA.}

\address{$^4$ Department of Physics and Astronomy, University of California,
Irvine, CA 92697-4575 USA}

\begin{abstract}
If the diffuse extragalactic gamma ray emission traces the large
scale structures of the universe, peculiar anisotropy patterns are
expected in the gamma ray sky. In particular, because of the
cutoff distance introduced by the absorption of 0.1-10 TeV photons
on the infrared/optical background, prominent correlations with
the local structures within a range of few hundreds Mpc should be
present. We provide detailed predictions of the signal based on
the PSCz map of the local universe. We also use mock N-body
catalogues complemented with the halo model of structures to study
some statistical features of the expected signatures. The results
are largely independent from cosmological details, and depend
mostly on the index of correlation (or bias) of the sources with
respect to the large scale distribution of galaxies. For instance,
the predicted signal in the case of a quadratic correlation (as it
may happen for a dark matter annihilation contribution to the
diffuse gamma flux) differs substantially from a linear
correlation case, providing a complementary tool to unveil the
nature of the sources of the diffuse gamma ray emission. The
chances of the present and future space and ground based
observatories to measure these features are discussed.

\end{abstract}
\pacs{95.85.Pw, 
98.70.Vc,    
95.35.+d 
}

\maketitle

\section{Introduction}
Gamma ray astronomy is a flourishing field in astroparticle
physics. Results have rapidly accumulated in the last decade or
two: after the break-through results of the EGRET satellite, a
whole series of Earth-based observatories have developed, both
using the imaging air Cherenkov telescopes (ACT like WHIPPLE,
HEGRA, CANGAROO, HESS, and MAGIC) or surveying the sky via
extensive air showers (EAS detectors, like TIBET, ARGO, and
MILAGRO). New projects using all these techniques (the ACTs
VERITAS, HESS II, MAGIC II, MACE, the EAS detector HAWC, the
satellites AGILE and GLAST) are on the way (for a recent review of
the field, see \cite{Ong:2006bz}).

At TeV energies, a few dozen sources have been detected mainly by
air Cherenkov experiments, most of which are high energy
counterparts of MeV-GeV sources in the EGRET catalog. The 0.1--10
TeV range represents one of the ``last'' photonic windows yet to
be explored at large distances. Starting from an energy of about
100 GeV (which we shall refer to as very-high energy, VHE), the
absorption of high energy photons onto the extragalactic
background light (EBL) via pair-production introduces an
energy-loss horizon of the order of a few hundreds Mpc or smaller,
well below the size of the observable Universe. This horizon drops
down to $\sim$10 kpc at PeV energies, basically precluding deep
space astronomy with photons above about 10 TeV. Far from being
only a limitation, this phenomenon also allows the use of
$\gamma$-ray astronomy to probe the EBL, which is otherwise
difficult to study directly.

Besides single sources, wide field of view instruments and
satellite-based observations are sensitive to diffuse $\gamma$-ray
emissions. A particularly interesting emission is the extragalactic
diffuse $\gamma$-ray background (in the following, cosmic gamma
background, or CGB). The CGB is a superposition of all unresolved
sources emitting $\gamma$-rays in the Universe and provides an
interesting signature of energetic phenomena over cosmological
time-scales. While a clear detection of this background has been
reported by the EGRET mission~\cite{Sreekumar:1997un}, its origin is
still uncertain, despite the fact that many models have been
proposed. The most likely contribution is the one from unresolved
blazars, i.e.~beamed population of active galactic
nuclei~\cite{Stecker:1996ma}, with (probably sub-leading) components
from ordinary galaxies~\cite{Pavlidou:2002va}, clusters of
galaxies~\cite{Gabici:2002fg}, and gamma ray
bursts~\cite{Totani:1998xc}. However, exotic possibilities like dark
matter annihilation have been proposed, that are compatible with
existing data and
constraints~\cite{Bergstrom:2001jj,Ullio:2002pj,Elsaesser:2004ck,Elsaesser:2004ap}.
It is extremely difficult to test such models as long as the only
observable is the energy spectrum. Recently, it was proposed to use
the peculiar small-scale anisotropy encoded in the MeV-GeV gamma sky
to probe dark matter~\cite{Ando:2005xg} or astrophysical
\cite{Zhang:2004tj,Ando:2006mt} contributions to the CGB. In this
work, we further study this topic, with particular emphasis on the
large scale anisotropy in the energy range 0.1-10 TeV. The lower
part of this range will be probed by the GLAST
telescope~\cite{Stecker:1999jh,McEnery:2004da}, while the energy
window above the TeV is in principle accessible to EAS detectors.
Different candidates to explain the CGB predict distinctive large
scale features, even when similar energy spectra are expected. This
is a consequence of the combined effect of a cutoff distance after
which VHE $\gamma$ can travel undamped to us, and of the anisotropic
distribution of matter in the local universe (i.e., within a few
hundred Mpc from us). A similar anisotropy pattern in the the
ultra-high energy cosmic ray sky was recently analyzed by some of
the authors~\cite{Cuoco:2005yd} (see also~\cite{Cuoco:2006dx}). The
main goal of this paper is to characterize various features of the
CGB which may be used for diagnostics at VHE.

We do not attempt here to derive the properties of the CGB from
specific astrophysical models, such as blazars (see
e.g.~\cite{Narumoto:2006qg}) or large-scale structure shocks
(see~\cite{Miniati:2002hs,Miniati:2003ep,Miniati:2007ke}).
Instead, we shall consider a phenomenological and parametric
approach, analyzing in detail the angular patterns vs. energy for
two representative diffuse background models: in the first case,
we assume that underlying gamma emitters correlate linearly with
overdensities in the large scale structure density field. In the
second case, we assume a quadratic correlation. Loosely, one may
consider the first case as representing unbiased, unresolved
astrophysical sources; the second one may be indicative of a
strongly biased astrophysical population of sources, or eventually
of dark matter annihilation emission\footnote{Of course, dark
matter emission would also have peculiar features in the energy
spectrum (e.g. departure from a power-law shape), associated with
the particle physics details. They may constitute an important
diagnostic tool for detection.}, provided it is not dominated by
substructures (see \cite{Ando:2006cr} for a discussion of this
point). Of course, the yet unknown sources of the CGB may have a
non-trivial bias and evolution with redshift. To a great extent,
additional information on this topic will be provided by the study
of the numerous point-like sources that GLAST is expected to
resolve. At present, our analysis should be considered as a
parametric approach to evaluate the sensitivity to ``effective"
matter tracing properties and bias of gamma ray sources.

The paper is structured as follows. The parametrization we use for
the CGB, based on EGRET data, is summarized in
Sec.~\ref{CGBmodel}. In Sec.~\ref{tracers} we introduce our
treatment of the cosmic large scale structure: a mock catalogue
derived from a dark matter N-body simulation and the 3D power
spectrum derived from the Halo Model of non-linear clustering
\cite{Seljak:2000gq,Peacock:2000qk,Ma:2000ik,Cooray:2002di}. With
these tools, we shall argue that in the VHE range, especially at
the largest angular scales, the predictions mostly depend on the
large scale structure in the local neighborhood of the Universe.
We shall then use the PSCz astronomical catalogue as tracer of the
real structures in the nearby universe, thus producing maps of the
VHE gamma sky (Sec.~\ref{skymap}). In Section \ref{detection} we
analyse the perspectives for detection of these features using the
forthcoming satellite GLAST, and for EAS observatories like
MILAGRO. In Sec.~\ref{conclusions} we summarize our findings, and
conclude. In ~\ref{prop} we provide some details on the
parametrization of the EBL, and on the method we use to propagate
gamma rays and calculate attenuation effects. \ref{ShotNoise}
reviews some statistical properties of a discrete poisson process
on the sphere, relevant for our estimates of the errors of the
multipole maps.

\section{The cosmic gamma background} \label{CGBmodel}
\begin{figure}[!tb]
\begin{center}
\begin{tabular}{c}
\epsfig{file=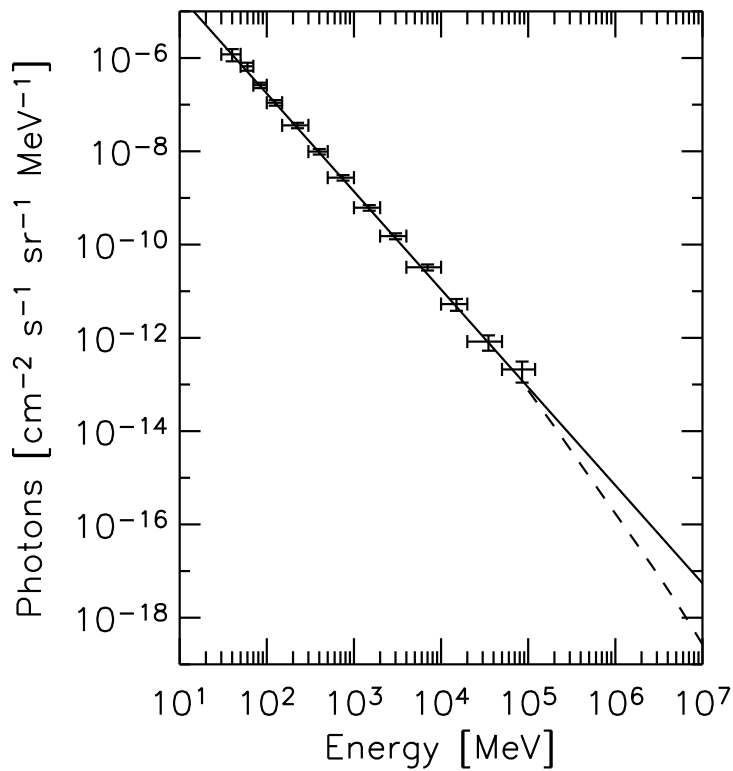,width=10cm}
\end{tabular}
\end{center}
\caption{EGRET spectrum from \cite{Sreekumar:1997un} and
extrapolation up to 10 TeV. The dashed line shows the expected
effect of the pair-production attenuation.\label{EGRETspectrum}}
\end{figure}

Experimentally, the CGB is the most difficult component of the
diffuse emission to study. Indeed, it is not correct to assume
that the isotropic component after extracting point-like sources
plus the galactic diffuse emission is entirely extragalactic: even
in the pole direction, the CGB is comparable to the Galactic
contribution. The deduced CGB thus depends on the adopted model of
the Galactic foreground. The analysis undertaken to derive the
spectrum of the CGB based on EGRET data provided the intensity
spectrum~\cite{Sreekumar:1997un}
\begin{equation}
I_\gamma(E)=k_0 \left(\frac{E}{0.451 {\rm GeV}}\right)
^{-2.10\pm0.03}\, , \label{spectrum98}
\end{equation}

valid from $E\sim\,$10 MeV to $E\sim\,$100 GeV, where
$k_0=(7.32\pm 0.34)\times 10^{-6} {\rm cm}^{-2}{\rm s}^{-1}{\rm
sr}^{-1}{\rm GeV}^{-1}$. Interestingly, it shows a spectral index
remarkably close to the average one of $\gamma$-ray blazars
detected by EGRET, $-2.1\pm 0.3$ \cite{Chiang95}. EGRET
experimental points and the best fit curve are shown in Fig.
\ref{EGRETspectrum}.

It is worthwhile to comment that the foreground subtraction
remains a delicate issue, as can be appreciated by the reanalysis
of the data performed in~\cite{Strong:2004ry}, based on a revised
model for the galactic propagation of cosmic rays. The deduced
extragalactic spectrum is significantly lowered with respect to
Eq.~(\ref{spectrum98}) at intermediate energies, while closer to
the original result of Eq.~(\ref{spectrum98}) at the lowest and
highest energy points. Since the removal of the isotropic galactic
emission from the diffuse gamma background is an open problem, and
we still lack a complete understanding of the sources of the CGB,
we shall simply base our following analysis on the extrapolation
of the spectrum of Eq.~(\ref{spectrum98}) by one to two orders of
magnitude. Actually, the anisotropy pattern in the CGB sky itself
may help in the foreground removal. Recently, it was proposed to
use the cosmological Compton-Getting effect (whose dipole
direction and amplitude are basically energy-independent) to
discriminate the truly extragalactic fraction from the galactic
foreground~\cite{Kachelriess:2006aq}.

When extrapolating the EGRET flux to higher  energies attenuation
effects must be taken into account. It is well known that the
propagation of photons in the EBL is a crucial issue for gamma
astronomy in the VHE range. Absorption of VHE photons through
pair-production with CMB, infrared or optical photons distorts an
initial source spectrum, in particular by steepening its high
energy tail. Astronomy of gamma ray emitters like blazars and
active galaxies requires then an accurate modelling of the photon
propagation and of the background frequency distribution. In
\ref{prop} we provide a detailed description of the model of the
EBL used, and of the treatment of absorption effects. For an
overview of the knowledge of the EBL we refer the reader to the
review~\cite{Hauser:2001xs}. The expression for the flux expected
at Earth in a generic cosmology is given by (see e.g.
\cite{Gao:1991rz,Ullio:2002pj})

\begin{equation} I(E_\gamma,\hat{n})\propto \int_0^\infty \d
z\,\frac{\rho^\alpha(z,\hat{n})\,g[E_\gamma(1+z)]\,e^{-\tau(E_\gamma,z)}}{H(z)\,(1+z)^{3}}\,
,\label{intcosmo}
\end{equation}
where we assume an universal spectrum for the source, $g(E)$,
$E_\gamma$ is the energy we observe today, $\rho(z,\hat{n})$ is the
matter density in the direction $\hat{n}$ at the redshift $z$, where
the sources are assumed to be distributed proportionally to
$\rho^\alpha$. The Hubble parameter is related to the present Hubble
expansion rate $H_0$ through the matter and the cosmological
constant energy density as $H(z)=
H_0\sqrt{\Omega_M(1+z)^3+\Omega_\Lambda}$. The quantity
$\tau(E_\gamma,z)$ is the optical depth of photons to absorptions
via pair production on the EBL (see \ref{prop}). For most of the
following considerations the normalization in the spectrum is
irrelevant. Where ever needed (e.g.~to estimate the statistics which
can be collected by a given experiment) we shall fix the
normalization of Eq. (\ref{intcosmo}) so that it matches the EGRET
fit of Eq. (\ref{spectrum98}) at 10 GeV.

For most of what follows it is important to realize that, by looking
at VHE, most of the dependence on the cosmology, the source
evolution, etc. in Eq. (\ref{intcosmo}) cancel out, independently of
the index of correlation with density, because of the cutoff at
$z\ll 1$ existing for VHE gammas. This important property
dramatically reduces the model-dependence of the following
considerations.
It is also worth commenting that the $\gamma$-rays of the CGB
constitute only a tiny fraction, $f_\gamma$, of the cosmic ray flux.
When compared with the flux of cosmic rays around the TeV ($I_{\rm
CR}=2.582\times 10^{-11}(E/{\rm TeV})^{-2.7}~{\rm cm}^{-2}\, {\rm
s}^{-1}\, {\rm sr}^{-1}\, {\rm MeV}^{-1}$ \cite{Aharonian:2001ft})
and neglecting gamma attenuation effects one gets the upper limit
\begin{equation}
f_\gamma\equiv \frac{I_\gamma}{I_{\rm CR}}\alt 2.7\times
10^{-5}\left(\frac{E}{{\rm TeV}}\right)^{0.6}\, .\label{gCRratio}
\end{equation}
Actually, the attenuation of $\gamma$'s on the EBL cuts the growth
of $f_\gamma$, which never exceeds 10$^{-5}$.
Note finally that our extrapolation is
consistent with existing observational bounds on $f_\gamma$ at the
TeV scale~\cite{Aharonian:2001ft}.

\section{Tracers of the large scale structure}\label{tracers}
Since the sources of the CGB are unknown (this is especially true
for its high energy component of interest here), to predict the
anisotropy pattern in the VHE sky we must start from some assumption
on the distribution of the sources. In the following, we shall
assume that the sources of the CGB follow the LSS distribution of
the matter. Starting from Eq. (\ref{intcosmo}), the integral flux
above the energy $E_{\rm cut}$ at Earth is simply written as
\begin{equation}
F(E_{\rm cut},\hat{n})\propto \int_0^\infty \d
z\,\frac{\rho^\alpha(z,\hat{n})}{H(z)\,(1+z)^{3}}W(E_{\rm
cut},z)\, ,\label{smoothedmap}
\end{equation}
where we have defined the window function as
\begin{equation}
W(E_{\rm cut},z)\equiv \,\int_{E_{\rm cut}}^{\infty}\d E\,g[E(1+z)]
\,e^{-\tau(E,z)}\, .\label{gammawindow}
\end{equation}
In the limit where the effective cutoff $z_c\ll 1$, the integrals
become almost independent on the cosmology, and the above
expressions simplify considerably to
\begin{eqnarray}
W(E_{\rm cut},z)&\simeq& \,\int_{E_{\rm cut}}^{\infty}\d E\,g(E)
\,e^{-\tau(E,z)}\, ,\label{Wnear}\\
F(E_{\rm cut},\hat{n})&\propto& \int \d
z\,\rho^\alpha(z,\hat{n})\,W(E_{\rm cut},z)\, ,\label{Fnear}
\end{eqnarray}
where $z$ is directly proportional to the distance, $r\simeq c\,
z/H_0.$ In particular we shall assume in the following the
power-law $g(E)\propto E^{-2.1}$, which on the theoretical side is
consistent with a Fermi-shock acceleration mechanism, and
observationally matches both the average spectral index of blazar
and the spectra of the CGB measured by EGRET, see Eq.
(\ref{spectrum98}).

The size of the horizon is roughly 300 Mpc $h^{-1}$ at 1 TeV (see
\ref{prop}) allowing safely the use of existing astronomical
catalogues like PSCz to predict the gamma signal. At lower energy,
statistical methods like N-body simulations and the Halo model of
structures are in principle required to complement existing
catalogues, which are not deep and/or wide enough. We shall
describe these tools in the next section and use them to derive
the expected statistical properties of the anisotropies at $E<1$
TeV. These analyses indicate that, at least at large angular
scales, most of the non-statistical information contained in the
surveys can still be retained, allowing one to use the exiting
catalogues for cross-correlation studies even at $E<1$ TeV.

{\bf N-Body Simulations} -- Our primary tool to compute the
statistical properties of the matter density distribution is a
N-body catalogue. This offers the advantage compared to
observations of being virtually free of most of the bias affecting
astronomical catalogues (no extinction regions, no selection
effects, no galaxy morphology and color bias, etc). On the other
hand, to obtain reliable results one should carefully consider the
types of bias introduced by the simulation, like the algorithm
used, the minimum scale resolved, and the physical content and
processes treated in the numerical experiment.

We have performed a set of pure dark matter N-body simulations with
a standard $\Lambda$CDM universe, a periodic box size of
$800~h^{-1}\, \textrm{Mpc}$, and resolutions of $512^3$ and $768^3$
particles using the GADGET2 code
\cite{Springel:2005mi,Springel:2000yr}. The initial conditions were
computed using second order Lagrangian perturbation theory
\cite{Crocce:2006ve}. Placing the observer at an arbitrary point in
the box, we have computed the density on logaritmically spaced
spherical shells. A smoothed density field $\rho(z,\hat{n})$ is
reconstructed with an adaptive algorithm whose smoothing length
varies depending on the local number density of dark matter
particles identical to the one used in the GADGET2 code. After
obtaining the window functions for different energy cuts, the sky
map of the integral gamma ray flux above the cut is calculated by
integration along radial lines of sight, considering a source
density proportional to $\rho$ or to $\rho^2$. We have checked that
the resolution used in the simulation is sufficient to suppress shot
noise from undersampling and other numerical artifacts, by comparing
the power spectrum extracted at different resolutions. The resulting
maps $F(E_{\rm cut},\hat{n})$ are obtained on a Healpix grid,
facilitating the use of standard tools from CMB physics to compute
the final power spectrum and multipole coefficients $C_l$
\cite{Gorski:2004by}. For the final analysis we used 2048 shells,
with $12\times1024^2$ pixels on each shell or a density grid with
almost $25\times10^9$ pixels. This makes it possible to reconstruct
the power spectrum reliably up to $l=3071$.

{\bf Halo Model} -- In order to check the results of the N-body
simulations, we compare the $C_l$'s obtained in the case of linear
correlations with the predictions using the Halo Model
\cite{Seljak:2000gq,Peacock:2000qk,Ma:2000ik}. This model is known
to provide a fast and efficient semi-analytical recipe for
describing the clustering of dark matter halos, and of their
evolution. Comprehensive reviews of the model can be found
in~\cite{Cooray:2002di,Smith:2002dz}. Assuming that the gamma
emitters follow the distribution of the halos, the angular power
spectrum of the adimensional gamma flux anisotropy $\left[F (E_{\rm
cut}, \hat{n}) - \left<F (E_{\rm cut}) \right> \right] / \left<F
(E_{\rm cut}) \right>$ can be calculated with the 3D matter power
spectrum $P(k)$ and its normalization provided by the model. In the
flat-sky limit and the Limber approximation, it simply reads
\begin{equation}\label{limber}
    C_l=\frac{1}{N_w^2} \int \frac{dr}{r^2} \; W^2(E_{\rm cut},r) \;
    P\left(k=\frac{l}{r},z(r)\right)\, ,
\end{equation}
where the pre-factor $N_w \equiv \int \d r \, W(E_{\rm cut},r)$
represents the contribution from the mean intensity $\langle
F(E_{\rm cut})\rangle$.

\begin{figure}[!tb]
\begin{center}
\begin{tabular}{c}
\epsfig{file=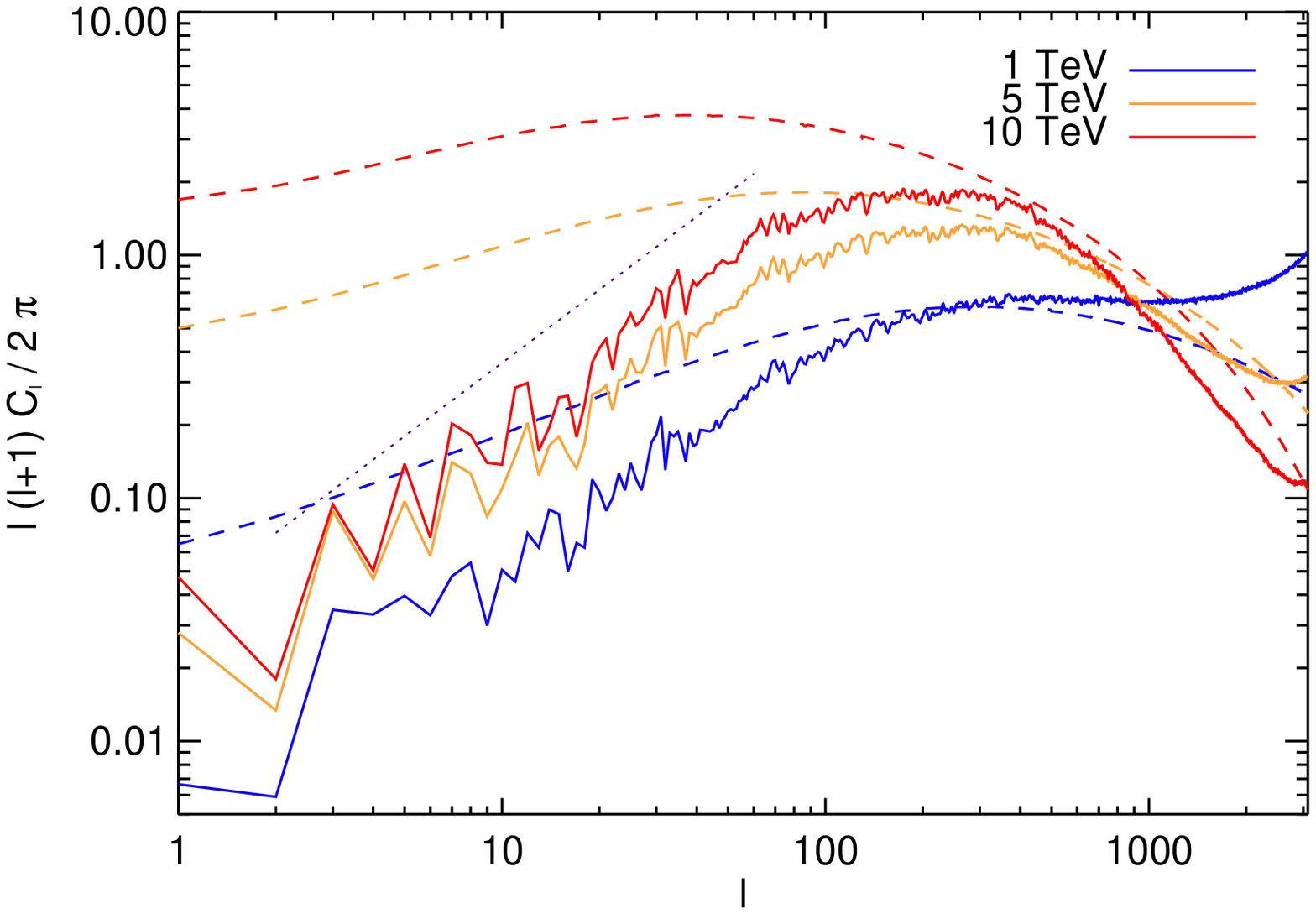,width=11cm} \\
\epsfig{file=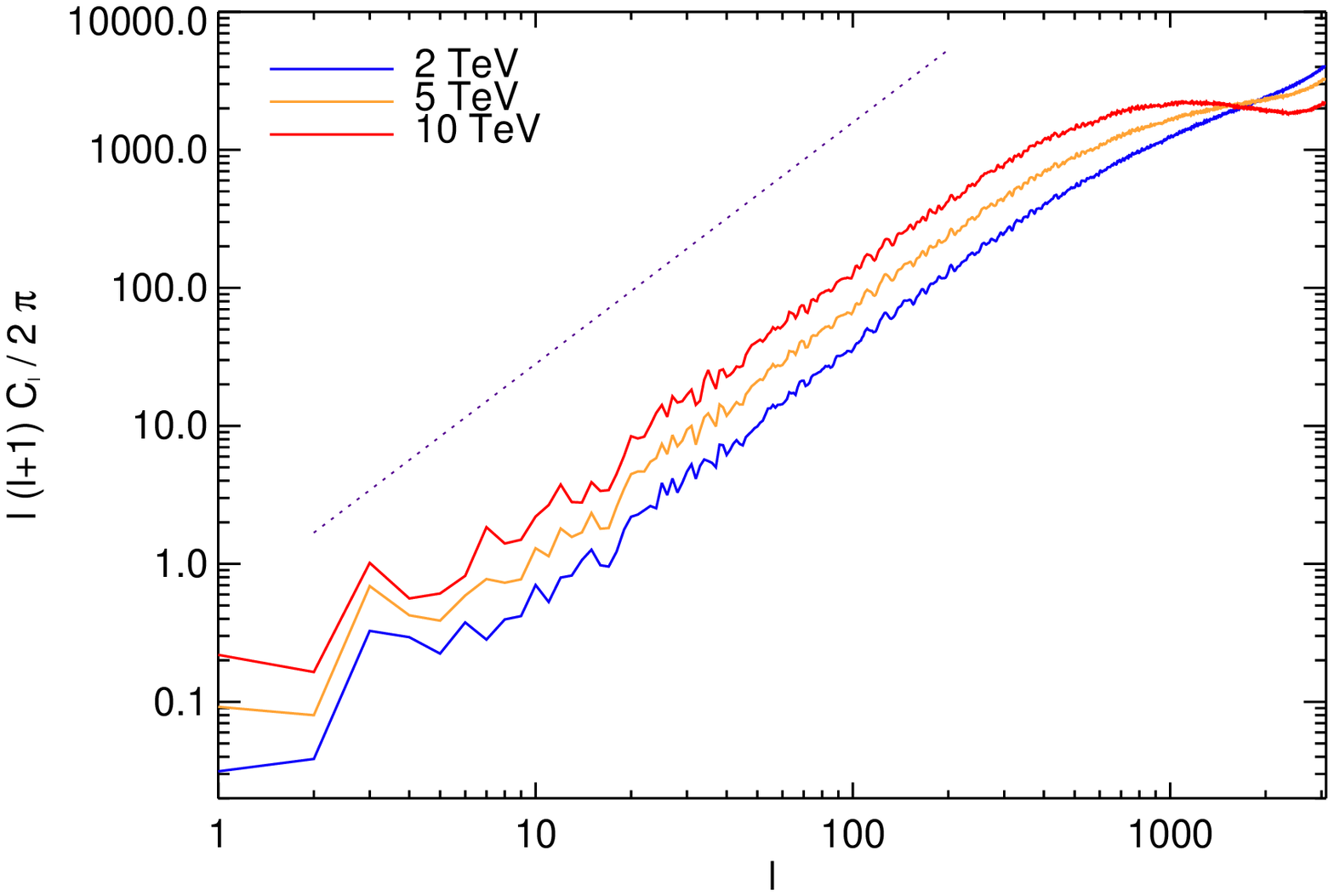,width=11cm}
\end{tabular}
\end{center}
\caption{In the top panel we show analytical angular spectra
(dashed lines) versus N-body spectra (solid lines) for energy cuts
$E_{\rm cut}=1,5,10$ TeV (from bottom to top), in the hypothesis
of linear correlation with matter. In the bottom panel we show
N-body spectra for a quadratic correlation. In both cases, the
dotted line shows the slope of the approximately linear rise of
the first multipoles ($l\ll 100$) of the power spectrum. The flux
from the N-body simulation is normalized by fixing the monopole:
$C_0 = 4\pi$. \label{AngularSpectra}}
\end{figure}

A comparison of the Halo Model and N-body spectra for
various energy cuts is shown in the
top panel of Fig.~\ref{AngularSpectra}.
To properly normalize the flux resulting from the N-body catalogue
we fix the monopole $C_0 = 4\pi$, where the mean flux is derived
by integrating Eq.~(\ref{smoothedmap}) over the average source
density found in the simulation. We see that the agreement is
pretty good at $100 \alt l\alt 1000$, while unsatisfactory at very
low and high $l$. The former disagreement depends on the breakdown
of the flat-sky (see e.g. Ref.~\cite{Hu:2000ee}) and the Limber
approximation, while the latter depends on the inadequacy of the
Halo Model to properly take into account non-linearities for
wavenumbers $k \agt 10~h\,{\rm Mpc}^{-1}$. The biggest
disagreement is at 10 TeV, for which the window function $W$ is of
width $\sim 100~h^{-1}\, {\rm Mpc}$ and the non linear scales
are most important. In the N-body catalogue we can only account
for the flux coming from distances up to $350~h^{-1}\,{\rm Mpc}$.
At relatively low energies ($E_{\rm cut} < 2$ TeV) there are
essential contributions to the flux on small scales/high $l$ from
distances further away than that, while negligible contributions
to the flux on large scales/low $l$. At small angular scales and
lower energies the simulation also suffers of residual shot noise,
seeable in the break in the slope at high $l$, that makes the
comparison biased.

In the lower panel, we present the angular spectrum for the
hypothesis of a quadratic correlation with density. Both in the
linear and the quadratic case, the three most interesting features
are (i) the power law slope at low scales, which is approximately
energy independent, as indicated in Fig.~\ref{AngularSpectra} by
the dotted purple line, (ii) the presence of a peak (deriving from
the peak of the 3D matter power spectrum), and (iii) the
increasing relative amplitude of the anisotropies with growing
energy cut. At relatively large scales, the anisotropies are
contributed by the near structures, which are therefore poorly
affected by the shrinking of the pair-production energy loss
horizon. The isotropic or small-scale varying component of the
flux has instead a significant contribution from far objects,
which are cutted away at high energies. The main difference
between the linear and the quadratic case is clearly in the
normalization and the slope of the spectrum. In the latter
scenario the intensity of the anisotropies is enhanced, and the
slope is steeper. The slope can be estimated even at large
scales/low l, and it is a signature of the source correlation with
the matter density accessible to observations. For a linear
density correlation we find $l(l+1)C_l \propto l$ while for a
squared density correlation we find $l(l+1)C_l \propto l^{1.75}$.
Also interesting is the prominent shift of the peak to very large
$l$ in the quadratic case, although it is unlikely that this
feature may be observed experimentally.

\begin{figure}[!tb]
\begin{center}
\begin{tabular}{c}
\epsfig{file=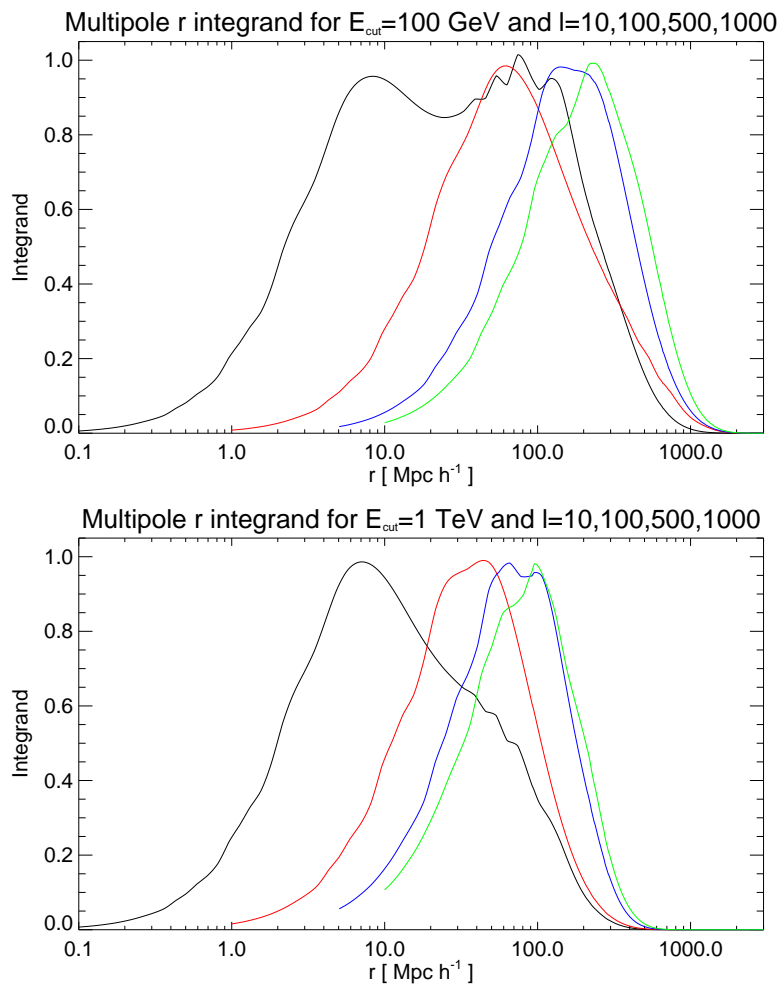,width=11cm}
\end{tabular}
\end{center}
\caption{Integrand contributions (linear in $\log r$) for the
multipoles $l=10,100,500,1000$ (respectively, black, red, blue,
green) and $E_{\rm cut}=100$ GeV, 1 TeV. Normalization is
arbitrary. \label{integrands100GeV5TeV}}
\end{figure}

\begin{figure}[!htb]
\begin{center}
\begin{tabular}{cc}
\epsfig{file=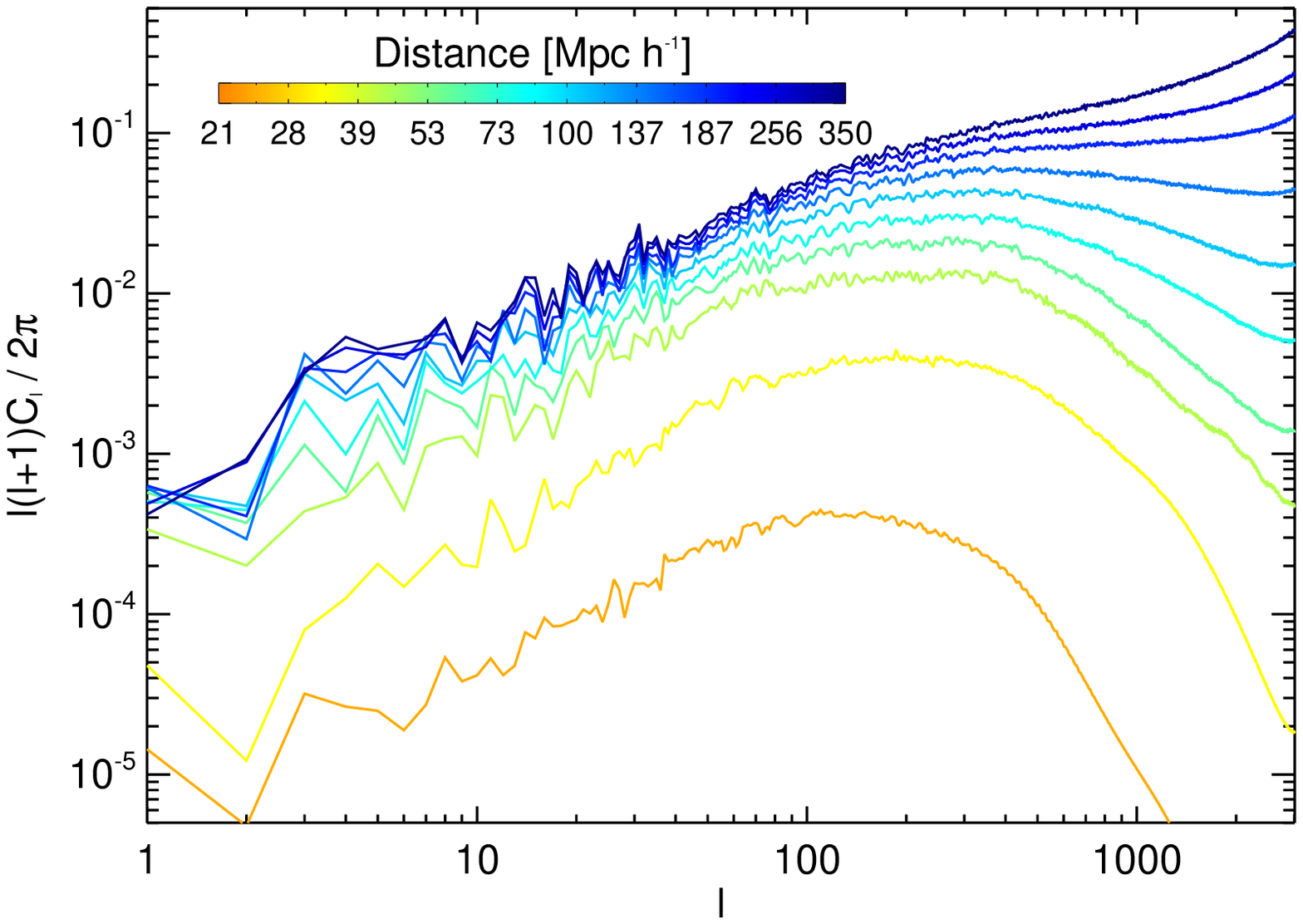,width=11cm} \\
\epsfig{file=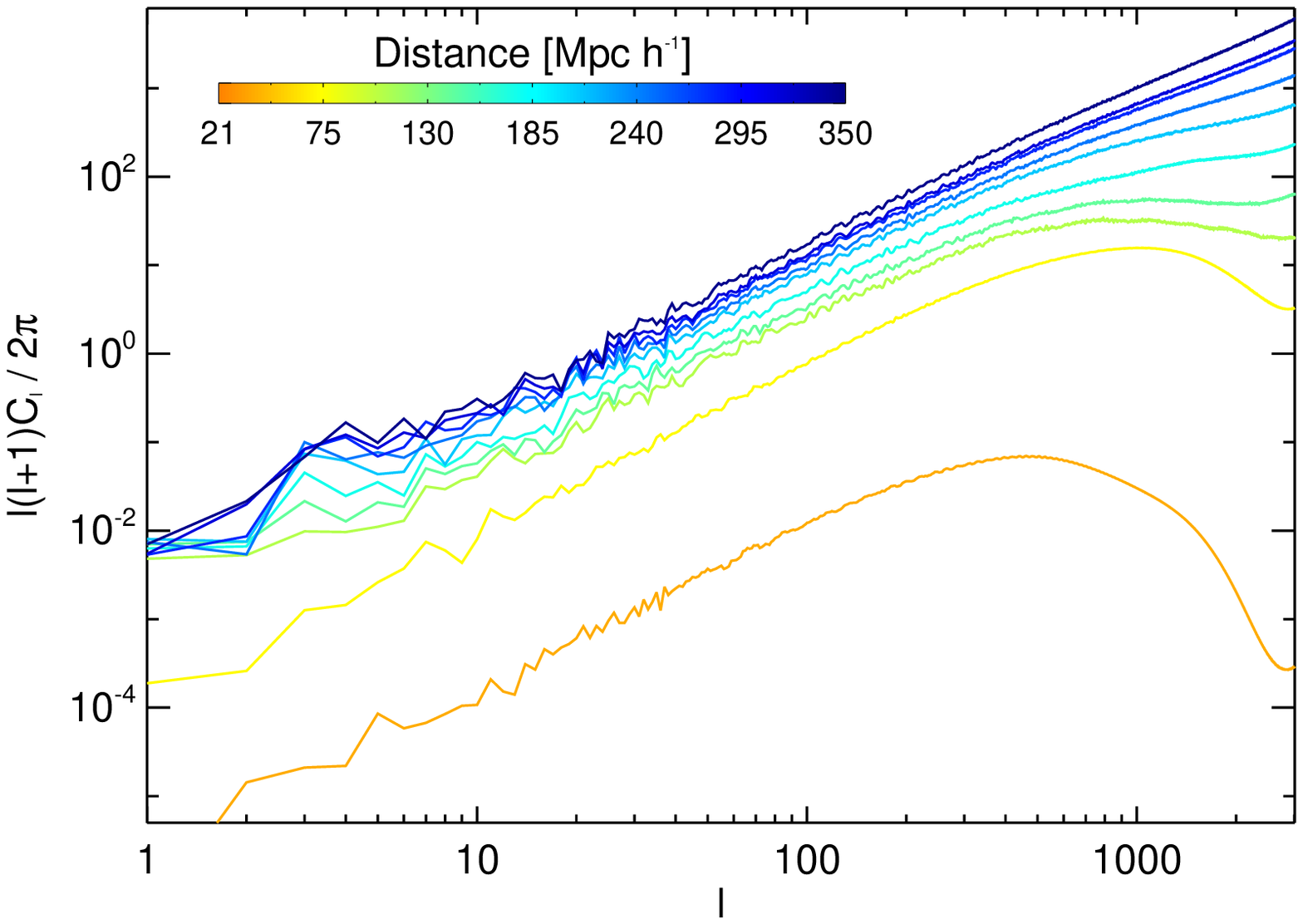,width=11cm}
\end{tabular}
\end{center}
\caption{100 GeV angular spectra when the N-body catalogue is used
only until a maximum distance $R$ as function of $R$. Shown are the
linear (top) and quadratic (bottom) density correlation case. \label{LinSqrIscp100GeV}}
\end{figure}

We turn now to study how shells at different distances $r$
contribute to the various multipoles. Using the predictions of the
Halo Model, in Fig.~\ref{integrands100GeV5TeV} we plot the integrand
$r^{-2} \; W^2(E_{\rm cut},r) \; P(k=l/r)$ versus $r$ for the two
energy cuts $E_{\rm cut}=$100 GeV and 1 TeV at different $l$'s. Note
that even for $E_{\rm cut}=$ 100 GeV, the $C_l$'s at $l=10-100$
mostly depend on the contribution from structures within a radius of
$\simeq 600~h^{-1}\, {\rm Mpc}$. Interestingly, the dominant
contribution to the first multipoles at $l=1-10$ comes from within a
distance of only $\sim 300~h^{-1}\, {\rm Mpc}$. Within such
distances astronomical catalogues with a large field of view exist
(e.g.~the 2MASS and PSCz catalogues), providing the actual
distribution of matter in the universe, and not only a statistical
information on the density field. Thus, one may aim at the study of
the pattern of the first few observables $a_{lm}$'s or,
equivalently, of the anisotropy map smoothed to the appropriate
resolution. Note that, the amplitude of matter fluctuation has a
measured maximum at a wavenumber (scale) corresponding about to $k
 = 0.02~h\, {\rm Mpc^{-1}}$ ($\lambda= 300~h^{-1}\, {\rm Mpc}$)
\cite{Tegmark:2003uf}. Beyond this distance the low $l$ multipoles
start to receive decreasing contributions and converge rapidly to
their asymptotic value. This is a well known expectation, although
only recently detailed study e.g. of the convergence of the dipole
has been performed \cite{Rowan-Robinson:1999mx}. Actually, as
illustrated in Fig.~\ref{integrands100GeV5TeV} the presence of the
window further accelerate the convergence of the low $l < 10$ sky
pattern. This point is further illustrated with the help of the
N-body simulation. In Fig.~\ref{LinSqrIscp100GeV} we plot the
cumulative contribution to the $C_l$'s in function of the maximum
distance used in the catalogue, i.e. derived from
Eq.~(\ref{Fnear}) when cutting the integral defining $F(E_{\rm
cut},\hat{n})$ at a distance $R$. The linear and quadratic cases
are shown, assuming $E_{\rm cut}$=100 GeV (at higher energies, the
effect is even more pronounced). Indeed, the simulation confirms
that the convergence to the asymptotic value is faster at low
$l$'s than at the higher ones. Effectively, for $l\alt 10-20$ it
is meaningful to look for correlations between the gamma sky and
known local structures well below the TeV scale. Of course, the
argument can be turned around: by masking structures in the nearby
sky from the map, the intrinsic low-$l$ multipoles would be
largely suppressed, and one may thus access more easily other
signatures as the cosmological Compton-Getting dipole
\cite{Kachelriess:2006aq}.

\section{Sky maps}\label{skymap}

\begin{figure*}[!tp]
\begin{center}
\begin{tabular}{cc}
\epsfig{file=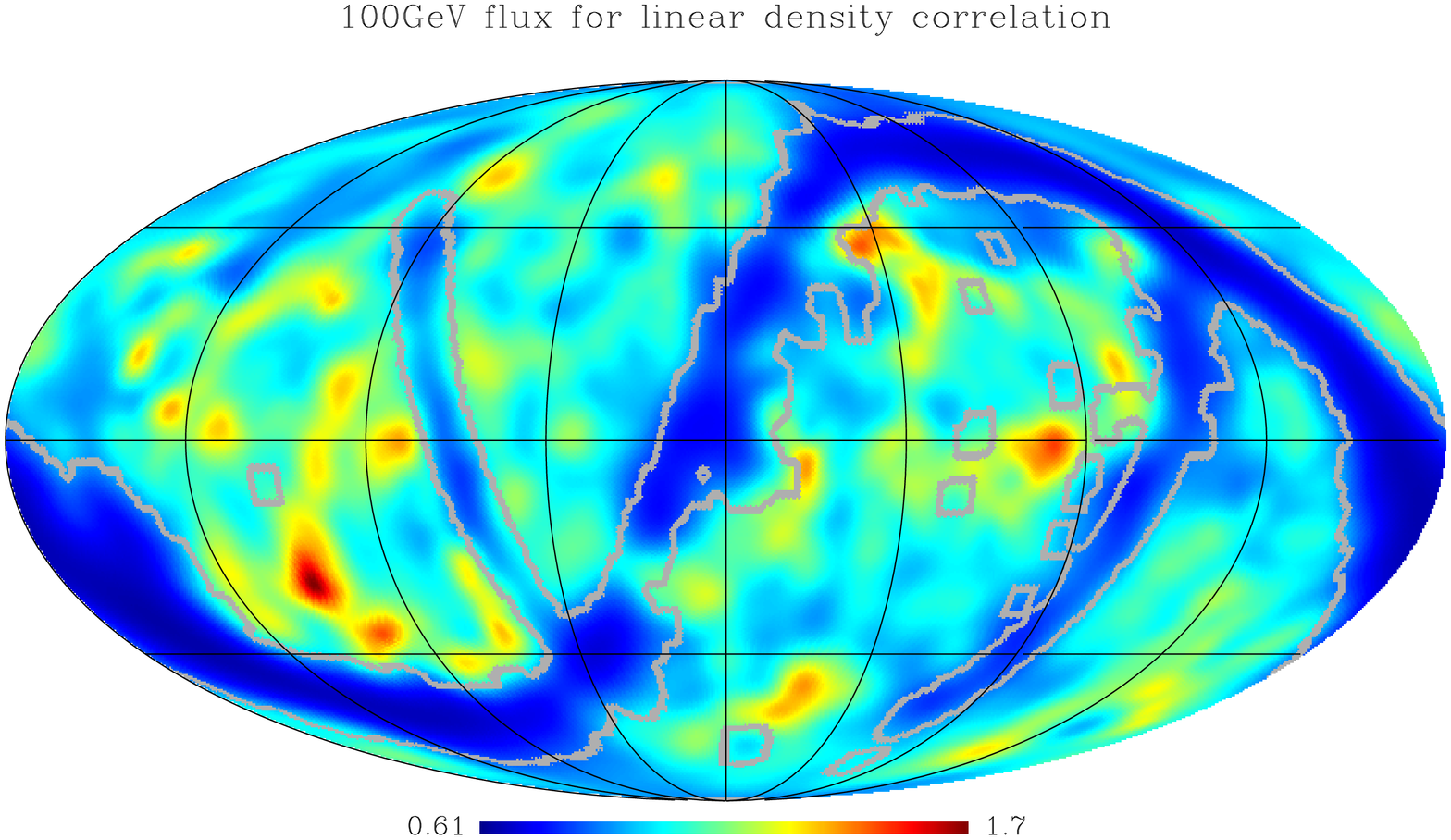,width=5cm,angle=90} & \hspace{1pc}
\epsfig{file=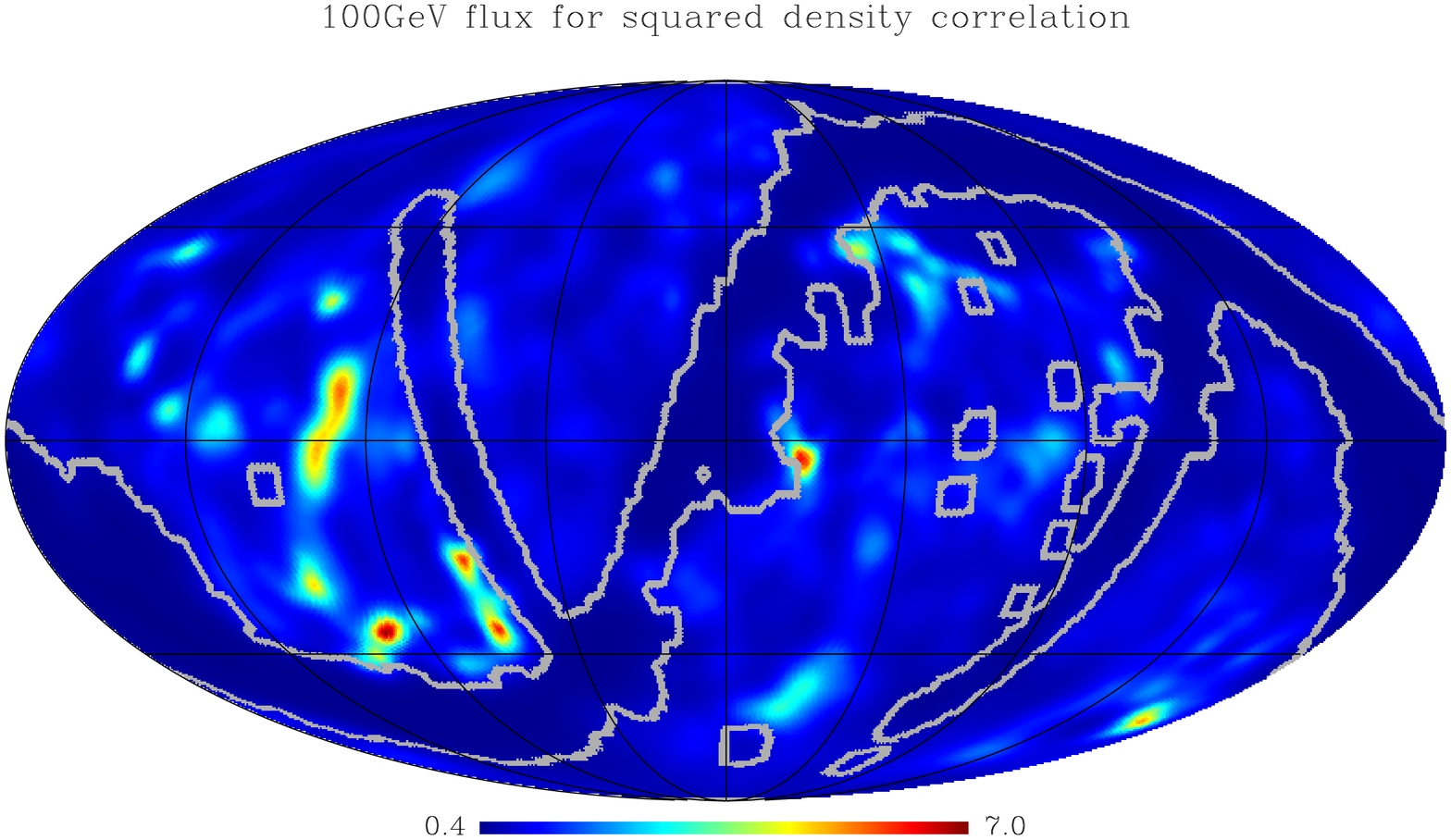,width=5cm,angle=90} \\
\epsfig{file=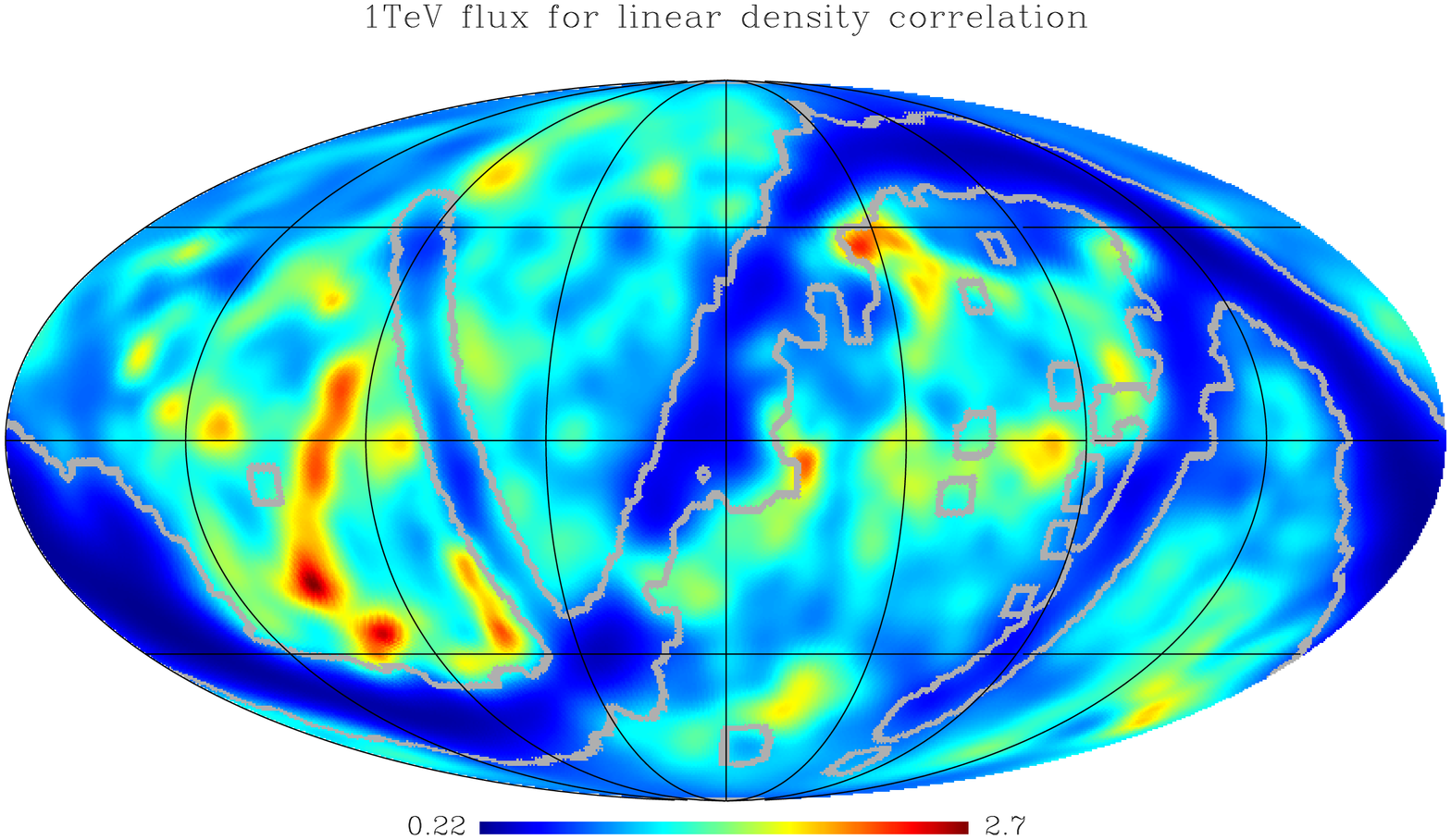,width=5cm,angle=90} & \hspace{1pc}
\epsfig{file=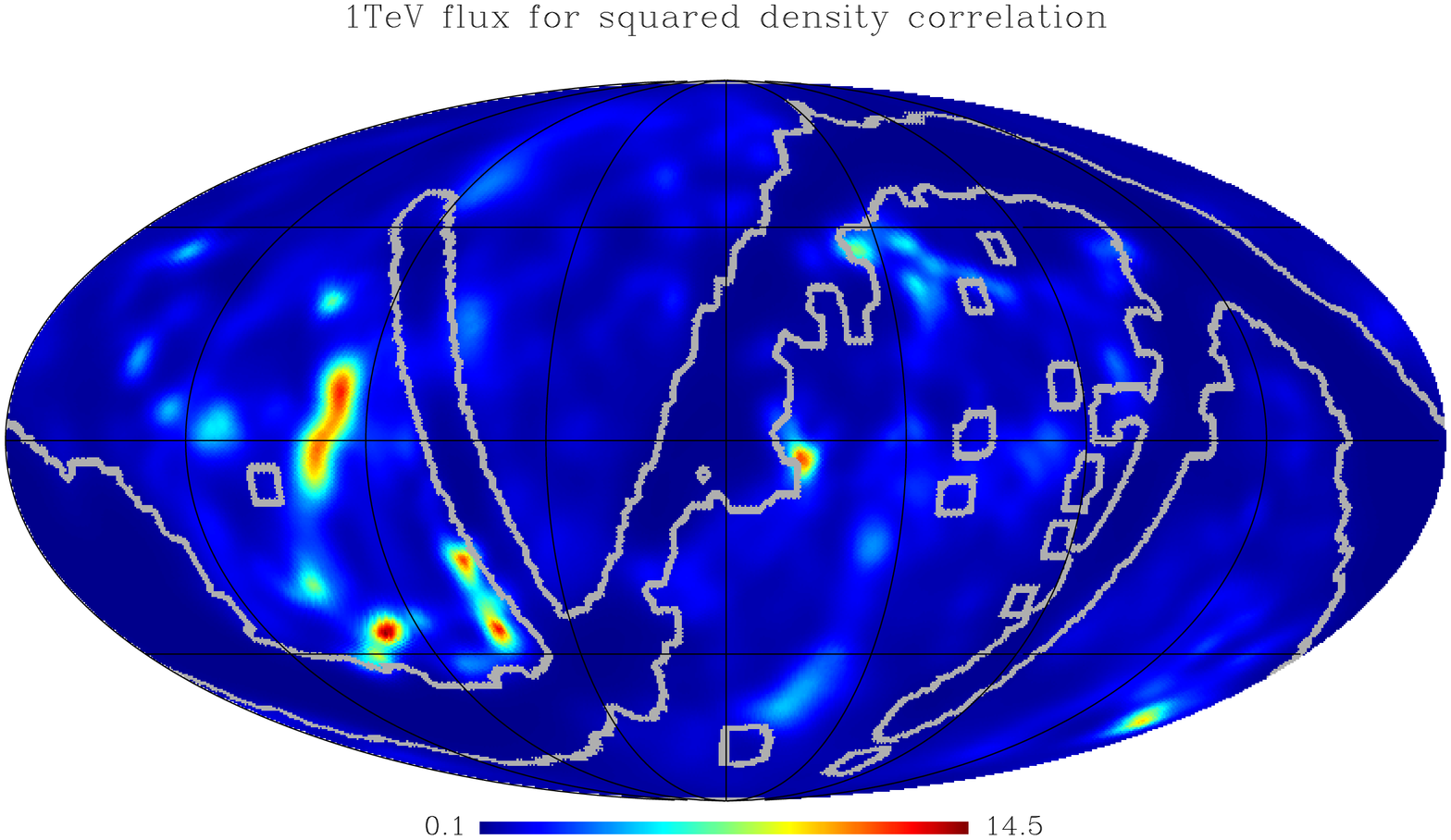,width=5cm,angle=90} \\
\end{tabular}
\end{center}
\caption{Equatorial density sky maps from the PSCz catalogue for the
linear (left) and quadratic (right) density correlation and for
$E_{\rm cut}=100$ GeV and 1 TeV. The color scale is linear and the
average flux outside the mask of the PSCz is normalized to 1 so that
to represent adimensional maps. The mask of the PSCz survey is
enclosed by the thick grey contour. \label{PSCzMaps}}
\end{figure*}

In the recent years, modern galaxy surveys like the Sloan Digital
Sky Survey~\cite{SDSS,Adelman-McCarthy:2005se} have greatly improved
our knowledge of the distribution of galaxies at large scales,
revealing a typical foam-like pattern of ``filaments and walls" of
galaxies around large cosmological voids. These very high quality
data are not well suited for our analysis due to the small fraction
of sky surveyed, while we need to perform comparisons with the large
sky field of view of GLAST and of EAS instruments. In this respect,
a fair compromise is offered by the IRAS PSCz
catalogue~\cite{saunders00a}.

The PSCz catalogue contains about 15,000 galaxies and related {\it
spectroscopic} redshifts with a well understood completeness
function out to $z \sim 0.1$. In the limit of uniform emission,
above $E=100\,$GeV the majority of the CGB flux is expected to
come from within this distance. The sky coverage of the catalogue
is about 84\%; the incompleteness is mainly due to the so called
zone of avoidance centered on the galactic plane and caused by the
galactic extinction and to a few, narrow stripes which were not
observed with enough sensitivity by the IRAS satellite. These
regions are excluded from our analysis with the use of the binary
mask available with the PSCz catalogue itself. However, the mask
region does not represent a major limitation for present analysis,
since the galactic emission at low galactic latitudes outshines
the CGB anyway, prohibiting any CGB analysis in this region.

We closely follow Ref.~\cite{Cuoco:2005yd} for the treatment of the
selection effects of the catalogue, parameterized via the selection
function $\phi(z)$. Differently from Ref.~\cite{Cuoco:2005yd}, we do
not sum source by source to obtain the final gamma sky map; instead,
we use the approach of constructing a smoothed density field
$\rho(z,\hat{n})$ through the same adaptive algorithm we employed in
the analysis of the N-body simulation. This method has the advantage
of efficiently suppressing the intrinsic catalogue shot noise at
high redshifts, thus allowing us to use the catalogue sources until
$z=0.1\simeq 300$ Mpc $h^{-1}$. We use Eq. (\ref{smoothedmap}),
while replacing $W(E_{\rm cut},z)\to W(E_{\rm cut},z)/\phi(z)$, to
take into account selection effects. To study the sensitivity to the
bias of the gamma sources with respect to the baryonic density, we
shall consider both values $\alpha=1$ and $\alpha=2$, i.e. a linear
and quadratic correlations. The latter is in particular expected for
dark-matter annihilation models \cite{Bertone:2004pz}.

In Fig.~\ref{PSCzMaps} we plot the resulting maps from the PSCz
catalogue in equatorial coordinates for the linear and quadratic
density correlation and for $E_{\rm cut}=$ 100 GeV and 1 TeV. Let's
discuss first the linear case. For the case of the map with $E_{\rm
cut}=100$ GeV, modulo the ``hole" due to the mask the pattern is
quite isotropic with some hot spots like from the Virgo and Perseus
Clusters. Other structures which appear are the Shapley
concentration and the Columba cluster (for a key of the local
cosmological structures see \cite{Cuoco:2005yd} or
\cite{Erdogdu:2006nd}). From the color scale can be seen that the
anisotropy could be also a factor of 2 in the hot spots, while is
generally of the order 10\% at larger angular scales as we will show
in the following with a multipole decomposition analysis. Given the
limited statistics of GLAST at high energies, the TeV map is of
interest especially for the EAS gamma detectors like MILAGRO. We see
in this case that the nearest structures, forming the Super-galactic
Plane, dominate. Of course, from the Northern emisphere (where all
the present or planned EAS instruments are located) only the upper
part of the map is visible. Here, the Virgo Cluster and the Perseus
cluster offer the strongest anisotropy.

For the quadratic case the change is quite evident: the effect of
the quadratic correlation is to give more power to the nearest
structures (the Virgo and Centaurus cluster) that, in fact, are
almost dominating the map with fluctuations exceeding even 10 times
the average and appearing almost like point sources. It is
instructive to look at the case of Shapley concentration at $z\simeq
0.04$ that gives an important contribution in all the linear cases
but almost disappears in the quadratic maps. The anisotropy in the
quadratic case is then much more pronounced than the linear case and
should be easily detectable.

\begin{center}
\begin{table}\footnotesize{
\begin{tabular}{|c|c|c|c|c|c|c|c|}
\hline Experiment & $A_{\rm eff}$ (cm$^{2}$) & $\Omega_{\rm fov}$
(sr) & $f_{\rm sky}$ & $DC$ &  $g_{\rm cut}$ &$h_{\rm cut}$ & range\\
\hline \hline GLAST\cite{GLAST} & $10^{4}$ & 2.4 & $f_{\rm m}$ &
$\sim$90\% &
 $\sim$1 & $\sim$.06 $f_\gamma$ & $E\alt$0.5-1 TeV\\
\hline MILAGRO\cite{MILAGRO} & $\agt 10^{7}$ & $\sim 2$ & $\alt
f_{\rm m}\times 2\pi$ & $>90\%$ &
 0.5  & 0.08 & $E\sim$1--20 TeV \\
\hline HAWC\cite{HAWC} & $\sim 10^{8.5}$ & $\sim$2 & $\alt f_{\rm
m}\times 2\pi$ & $>90\%$
& 0.5  & 0.08 & $E\sim$0.3--10 TeV \\
 \hline
\end{tabular}}
\caption{\label{Instrumental} The characteristics of the experiments
considered in our estimates. The fraction of the sky observable by a
given experiment $f_{\rm sky}$ is needed for our estimate of the
errors, see \ref{ShotNoise}.}
\end{table}
\end{center}
%

\section{The potential of forthcoming instruments}\label{detection}

In order to estimate the chances of detection of the structures
previously described by the current or next generation of
instruments, one may proceed as follows. First, given the specifics
of an experiment (in particular its field of view, effective area
and background rejection capability) one calculates the expected
number of events and of misidentified cosmic ray background events
$(N_\gamma,N_{\rm CR})$ under different assumptions for the $E_{\rm
cut}$ and the EBL. The events falling in the mask region must be
subtracted from the $(N_\gamma,N_{\rm CR})$ data used for the
analysis, so that the incomplete sky coverage is taken into account.
One then generates $\cal{N}$ samples of $N_\gamma$ events from the
VHE $\gamma$ map and $N_{\rm CR}$ from an isotropic one. These
$\cal{N}$ realizations may then be used to perform quantitative
statements, like for example the confidence level with which a given
instrument is expected to distinguish between the structured sky we
predict and an isotropic one, or between linear and quadratic
correlation scenarios, or between scenarios characterized by a
different model of CIB. This would be essentially the generalization
of the approach followed in \cite{Cuoco:2005yd}. Here, however, for
the sake of clarity we prefer to develop a simplified analysis in
terms of the multipole coefficients, $a_{lm}$, with an analytic
estimate of the expected errors due to shot noise effects (see
\ref{ShotNoise}). Performing full Monte Carlo analyses for a few
relevant cases we checked that the errors thus estimated basically
agree with the ones correctly calculated. In this approach, first we
fit the sky maps obtained outside the mask via the harmonic
expansion
\begin{equation}
F_{10}(E_{\rm cut},\hat{n})=
\sum_{lm}^{l_{\rm max}=10}
 a_{lm} Y_{lm}(\hat{n})\, .
\end{equation}
As a second step, by knowing the statistics and the characteristics
of the detector one can estimate the expected errors on the $a_{lm}$
and eventually on the derived coefficients $C_l$'s (see
\ref{ShotNoise} for details).

We limit the fit to $l\leq l_{\rm max}=10$ for several reasons:
(i) we have shown that the large scales are mainly sensitive to
the local universe, for which our predictions are robust and
deterministic; (ii) the lower the $l$, the higher the signal to
noise ratio is, which increases the chances of detection (see
below); (iii) the incomplete sky coverage due to the mask is
affecting our results, and should be taken into account; however,
as long as we restrict the analysis to the large structures
(compared to the size of the mask cut) the bias is small.
Quantitatively, the method is robust against variations in $l_{\rm
max}$ as long as the related angular scale $\theta_{\rm
min}=\pi/l_{\rm max}$ is greater than the typical angular
extension of the mask cut. More importantly, the purpose of the
harmonic analysis is to reliably assess the sensitivity of a given
experiment; to this aim a small bias in the $a_{lm}$'s is
acceptable (we do not want to estimate the true $a_{lm}$'s, but
the detectability of them, even of the biased ones). Of course, in
the analysis of the real data it will be preferable to directly
look for cross correlations with the full maps like the ones in
Fig.~\ref{PSCzMaps}, whose electronic version is available from
the authors upon request. Note that a naive a cross-correlation
between the gamma sky and the LSS catalogue would not take into
account the relevant window function effects we have highlighted
in this work.

\begin{figure*}[!tp]
\begin{center}
\begin{tabular}{cc}
\epsfig{file=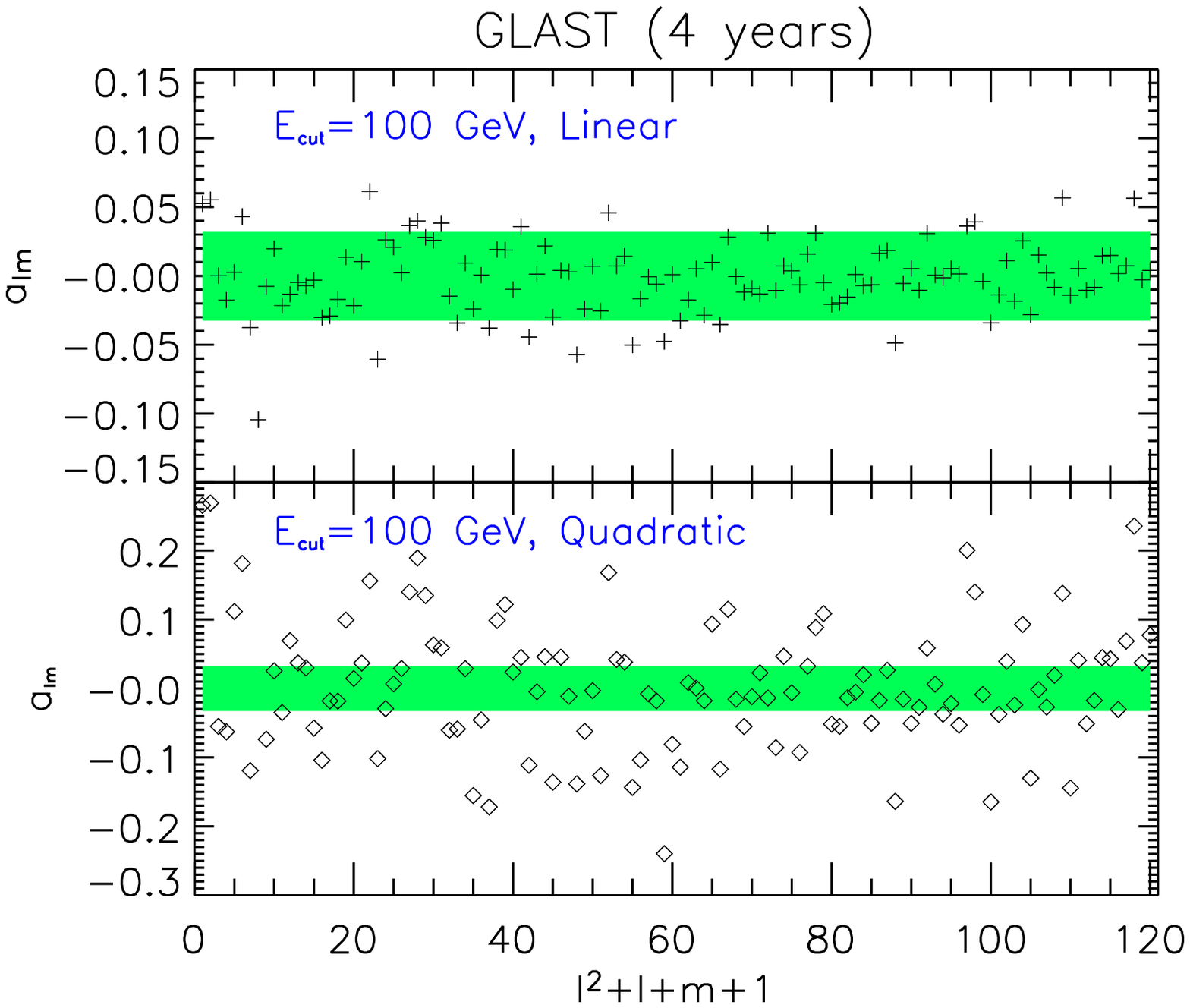,width=8cm,angle=0} & \hspace{0pc}
 \epsfig{file=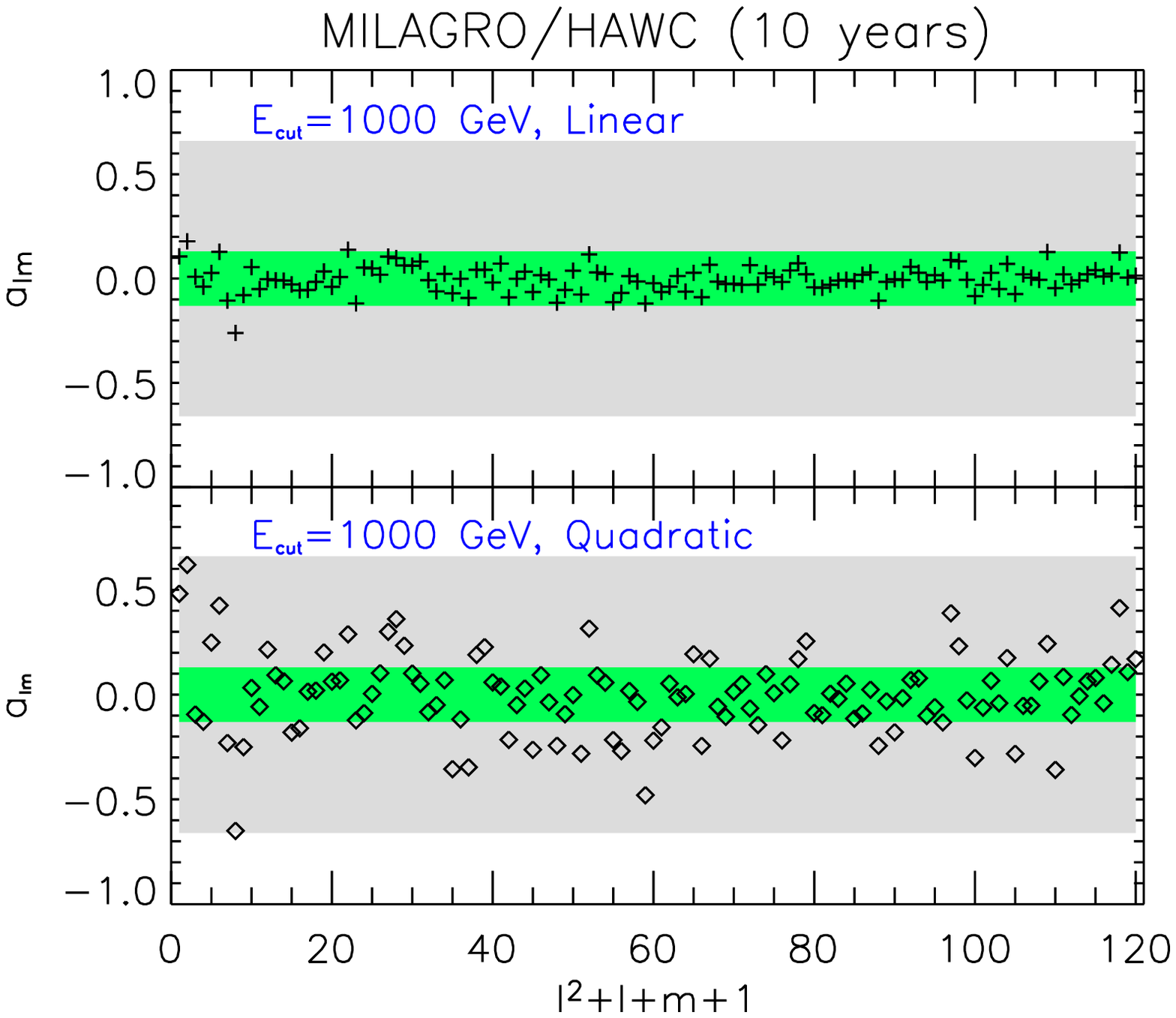,width=8cm,angle=0}
\end{tabular}
\end{center}
\caption{The coefficients $a_{lm}$ up to $l_{\rm max}=10$
calculated from the PSCz gamma maps of Fig. \ref{PSCzMaps}. The
shaded band shows the 1-$\sigma$ shot noise error given by Eq.
(\ref{sigmashotnoiselevel});  in the right panel the inner shaded
region refers to HAWC, the outer one to MILAGRO. We report the
predictions for both the linear and quadratic
cases.\label{PSCzAlms}}
\end{figure*}

In the following, we shall consider two kinds of instruments:
Satellite-based missions, in particular GLAST \cite{GLAST} (to be
launched in the Fall 2007), and extensive air shower experiments
like TIBET \cite{TIBET}, ARGO \cite{ARGO} (the assembly of which
started in October 2000 and is presently being completed), or
MILAGRO \cite{MILAGRO}. The latter experiment is taking science
data since 2001, and since 2004 in its outrigger hardware upgrade.
In the summer 2007, a transition of MILAGRO to the miniHAWC array
is expected to start; this is the first step towards the next
generation EAS planned observatories like HAWC \cite{HAWC}.
Unfortunately, ACT instruments like HESS or MAGIC are not well
suited for such kind of searches, given the small field of view,
the low duty cycle, and the relatively high impact of the
variability of the instrumental and atmospheric conditions on the
rate of diffuse signals.

Denoting by $I_\gamma(E)$ the extrapolated EGRET flux which takes
into account attenuations (see Section \ref{CGBmodel}), one can
estimate the number of events, $N_\gamma$, above the energy
$E_\gamma$ to be collected during the time $t$ as
\begin{equation}
N_\gamma = t\cdot g_{\rm cut}\cdot DC\cdot \Omega_{\rm fov}\cdot
f_{\rm m}\int_{E_\gamma}^\infty{\rm d}E\, A_{\rm eff}(E) I_\gamma(E)
\,, \label{Ngamma}
\end{equation}
where: $DC$ is the duty-cycle of the instrument; $\Omega_{\rm
fov}$ is the solid angle of the field of view; $f_{\rm m}<1$ is
the useful fraction of the sky due to the presence of the galactic
mask; $g_{\rm cut}$ is the fraction of $\gamma$'s passing the
actual cuts; $A_{\rm eff}(E)$ is the effective collecting area of
the instrument (averaged over the field of view of the
instrument). In the following, we shall assume $f_{\rm m}=0.84$
due to the mask in the PSCz catalogue, but note that future
redshift catalogues like 2MRS will have higher $f_{\rm m}$.
Eq.~(\ref{Ngamma}) assumes a quasi-isotropic $\gamma$ sky, which
may be violated to some extent at the multi-TeV energies of
interest for EAS detectors. Even in this case, right ascension
anisotropies would not affect the estimate, and only large
latitude anisotropies might affect $N_\gamma$ by a factor of
${\cal O}$(1). This is acceptable enough since we shall only
perform a parametric study of the performances of an EAS
observatory. Analogously, the CR background can be estimated as
\begin{equation}
N_{\rm CR} =t\cdot h_{\rm cut}\cdot DC\cdot \Omega_{\rm fov}\cdot
f_{\rm m} \int_{E_\gamma}^\infty{\rm d}E\, A_{\rm eff}(E)I_{\rm
CR}(E),
\end{equation}
where now $h_{\rm cut}$ is the fraction of hadrons passing the
cuts. Note that we consider the same area for CRs as for
$\gamma$'s, although a differential performance of the instrument
may be taken into account by properly rescaling the factor $h_{\rm
cut}$. The typical parameters we shall use are taken from existing
literature, and reported in Tab. \ref{Instrumental}. Note that
GLAST is expected to have an excellent background identification,
so that only cosmic rays in the amount of $\sim 6\%$ of the gamma
flux pass the cuts. On the other hand, EAS experiments have a poor
rejection capability (some of them like TIBET \cite{TIBET} have
none), which increases typically the gamma content of the diffuse
flux by no more than one order of magnitude. Therefore one should
keep in mind that even after gamma/hadron separation, the
anisotropies of the gamma sky have to be identified against a {\it
quasi}-isotropic background which is $\sim 10^{4}$ larger than the
gamma flux. The ultimate limitation in detecting anisotropies in
the gamma sky with EAS observatories is expected to come from the
understanding of the intrinsic anisotropy in the CR background. We
shall come back to this point in the conclusions.

\begin{figure*}[!tp]
\begin{center}
\vspace{2pc}
\begin{tabular}{cc}
\epsfig{file=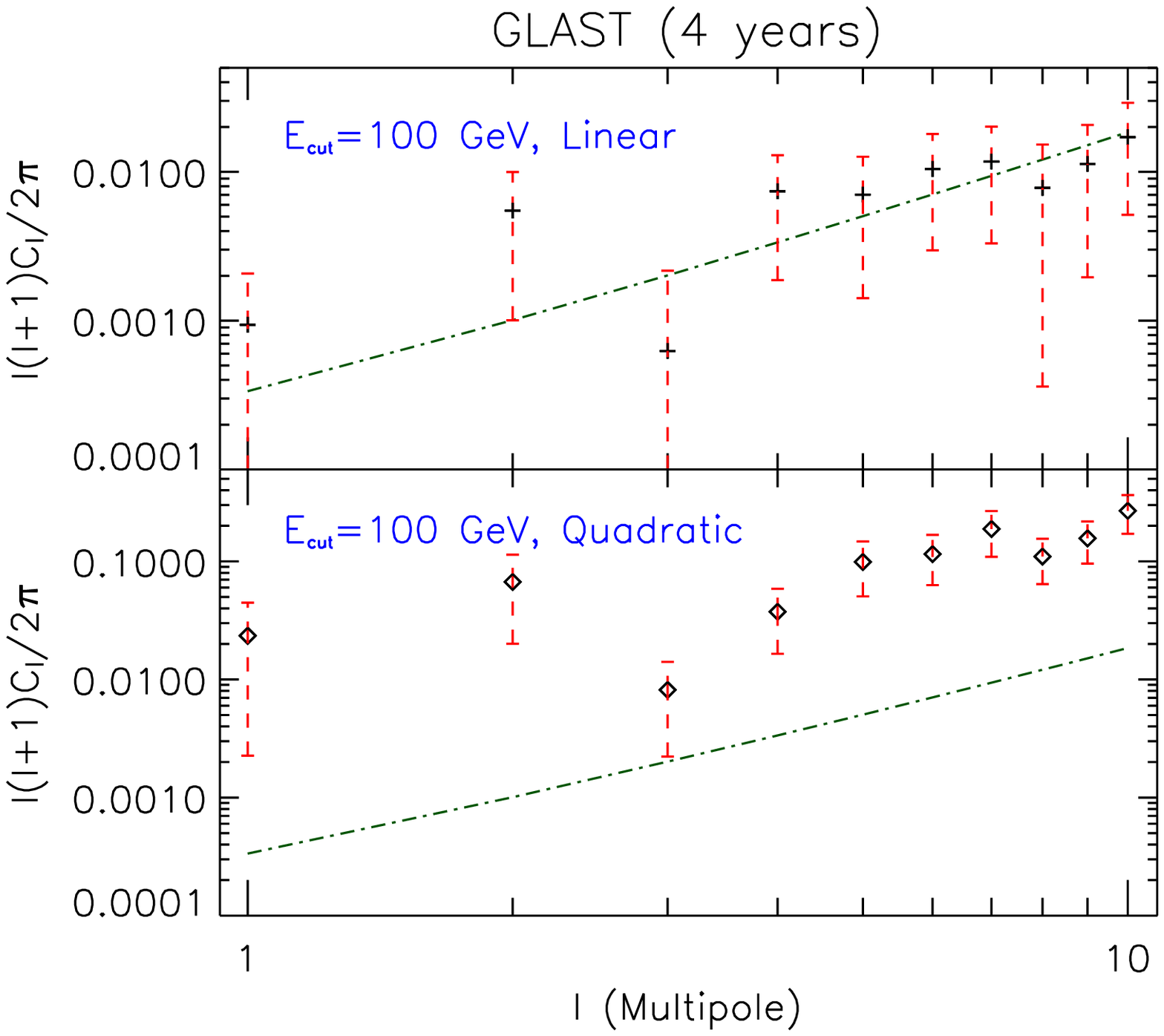,width=8cm,angle=0} & \hspace{0pc}
\epsfig{file=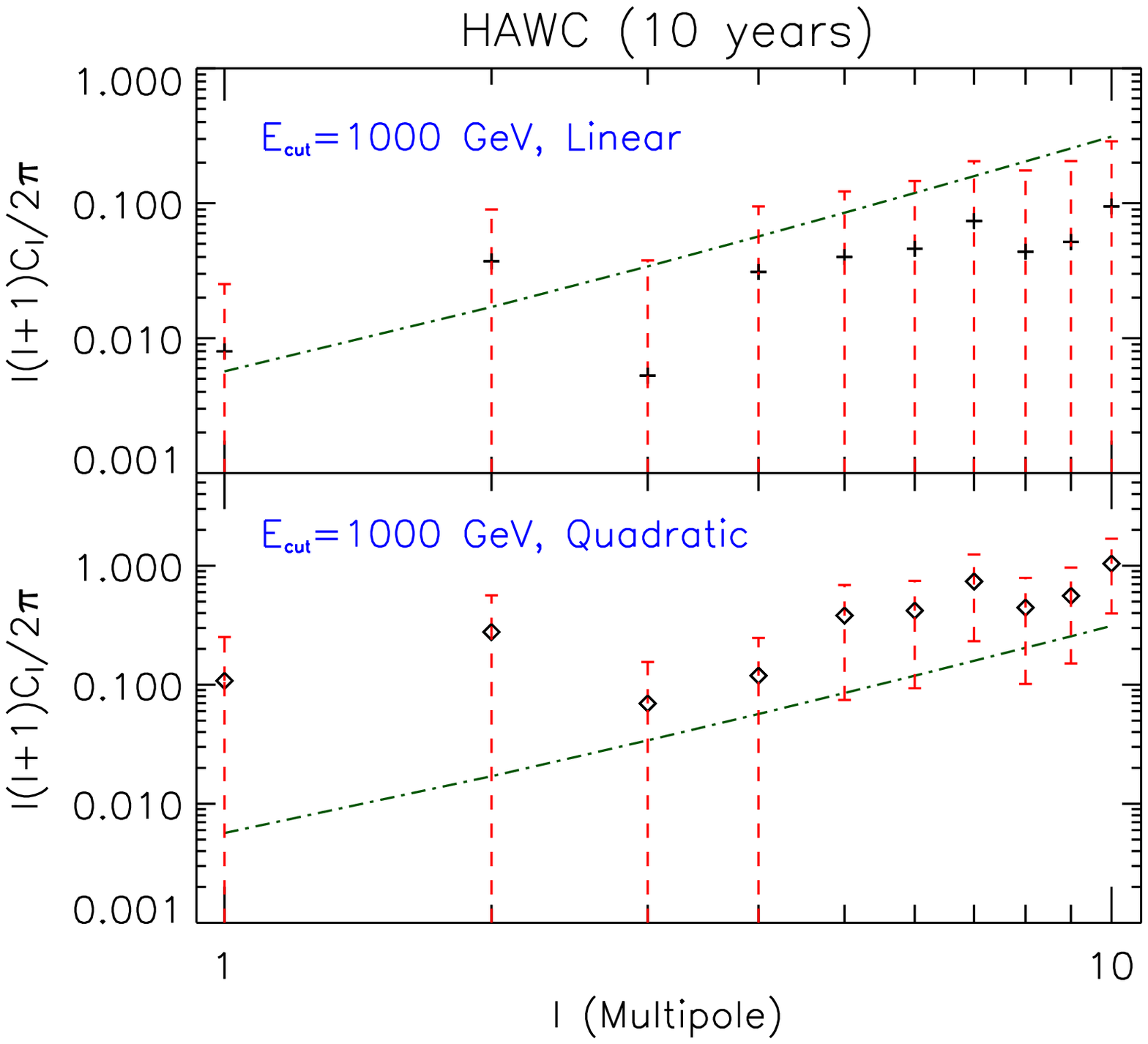,width=8cm,angle=0}\\
\end{tabular}
\end{center}
\caption{The coefficients $C_l$ up to $l_{\rm max}=10$ calculated
from the PSCz gamma maps of Fig. \ref{PSCzMaps}.  We shown the
level of shot noise [see Eq. (\ref{shotnoiselevel})] expected from
four years of GLAST and one decade of HAWC. We report the
predictions for both the linear and quadratic
cases.\label{PSCzCls}}
\end{figure*}

In Fig. \ref{PSCzAlms} we report the coefficients $a_{lm}$'s up to
$l_{\rm max}=10$ calculated from the PSCz gamma maps of
Fig.\ref{PSCzMaps}, with the errors estimated according to what
reported in \ref{ShotNoise}. The GLAST mission is expected to last
5+5 years (so we plot a realistic 4 years exposure), while EAS
instruments have longer run times, so we plot the expectations for a
decade of collecting time by MILAGRO, or by the proposed project
HAWC. GLAST should be able to detect some structures above 100 GeV
at the 2$\sigma$ level, even if the correlation with matter density
is only linear. For a quadratic correlation one expects a more
robust detection, and possibly even hints for anisotropies at higher
energies\footnote{Our estimate does not include the fraction of the
CGB measured by EGRET which may be resolved by GLAST. If this is
removed, our predictions should be rescaled accordingly.}. On the
contrary, instruments like MILAGRO may find hints of structures (at
the 1 $\sigma$ level, see gray band in the right panel of Fig.
\ref{PSCzAlms}) only if correlations are quadratic or in any case
strongly biased with overdensities. As a technical remark, note that
the performances of MILAGRO above the TeV were estimated by using an
effective area $A_{\rm{eff}}(E_{\gamma})\simeq 10^{7.1}\,$cm$^2$
(see e.g. \cite{HAWC}). A proper treatment should take into account
the energy-dependence of the area, and calculate the expected sky
maps for MILAGRO weighting accordingly the integral maps. Given the
limited chances of this instrument to detect the features we have
described, we consider this simple estimate sufficient to illustrate
our point. It is worth to stress that for an EAS detector the error
on the $a_{lm}$'s scales as $\sqrt{N_{\rm CR}}/N_\gamma$. Therefore
the reduction of the shot-noise error goes like $(t\cdot A_{\rm
eff})^{-1/2}$ (both $N_{\rm CR}$ and $N_\gamma$ grow linearly with
$t\cdot A_{\rm eff}$), or equivalently as $\sqrt{h_{\rm cut}}/g_{\rm
cut}$: improving the exposure is equally important as improving the
gamma/hadron separation capability. A simple inspection of Fig.
\ref{PSCzAlms} reveals that for a realistic detection of the
features in the VHE sky one would need the improvement in effective
area planned to be reached by instruments like HAWC \cite{HAWC} (see
inner green band in the right panel of Fig. \ref{PSCzAlms}). An
instrument like ARGO \cite{ARGO} is expected to have performances in
between MILAGRO and HAWC, and may have some chance especially if a
significant improvement in hadron rejection can be made. Also, note
that due to their altitude HAWC and ARGO have a significant
acceptance of sub-TeV events. While the gamma/hadron separation is
less efficient at lower energies, the higher statistics may help in
revealing these structures.

In Fig. \ref{PSCzCls} we plot the expectations for GLAST and HAWC in
terms of the $C_l$ coefficients. At the large scales we are focusing
on cosmic variance makes any detection of the $C_l$'s challenging
even when the corresponding $a_{lm}$'s are easily detectable. This
proves the importance of the deterministic nature of the expected
anisotropies in the flux.

\begin{figure*}[!tp]
\begin{center}
\vspace{2pc}
\begin{tabular}{c}
\epsfig{file=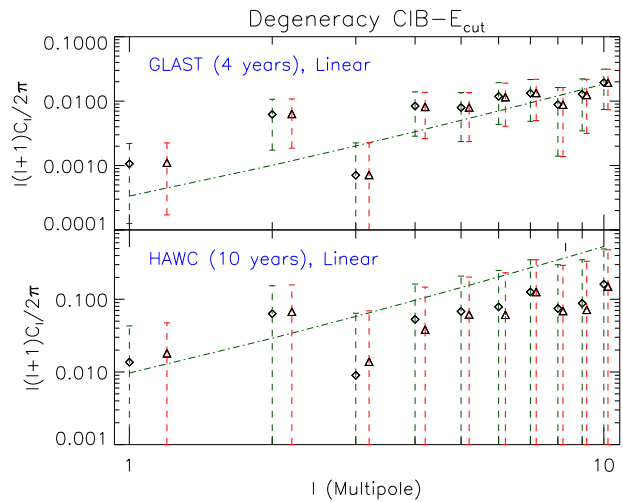,width=8cm,angle=0}\\
\end{tabular}
\end{center}
\caption{Upper panel: the coefficients $C_l$ up to $l_{\rm
max}=10$ calculated from the PSCz gamma maps of Fig.
\ref{PSCzMaps} as in Fig. \ref{PSCzCls}, assuming the fiducial
value of the CIB optical depth and an energy cut $E_{\rm{cut}}=$
120 GeV (diamond symbols, black errors bars) and a CIB value twice
the fiducial one (see Appendix A) and $E_{\rm{cut}}=$ 100 GeV
(triangle symbols, red errors bars). The value $E_{\rm{cut}}=$ 120
GeV is chosen so that the noise level (dot-dashed line) is the
same for the the two CIB values used. Lower panel: same as above
for $E_{\rm{cut}}=$ 2 TeV (fiducial CIB value, diamond symbols,
black errors bars) and $E_{\rm{cut}}=$ 1 TeV (twice fiducial CIB
value, triangle symbols, red errors bars.) \label{twicetau}}
\end{figure*}

Since our predictions are shaped by the nearby universe, for a
fixed background the absolute value of the detected anisotropy in
principle measures the index of the correlation of gamma sources
with respect to the matter, as clearly shown by the comparison of
the top and lower panels in both  Fig. \ref{PSCzAlms} and Fig.
\ref{PSCzCls}. On the other hand, a degeneracy exists with the
intensity of the infrared background.While an increase in the CIB
results in an increased absorption and then in a lower statistics,
at the same time the horizon for gammas is shrinking and the
intensity of the anisotropy increases too so that the signal to
noise ratio remain almost unchanged. In Fig. \ref{twicetau} we
compare the two spectra resulting from an increase in the energy
cut and an increase in optical depth in order to maintain the same
collected statistics or, equivalently, the same noise level $C_N$.
Indeed, at low energies ($E \simeq 100$ GeV) the change in optical
depth only slightly reduces the statistics so that the degeneracy
is almost perfect. At higher energies ($E \simeq 1$ TeV) the
change in statistics is more pronounced and the degeneracy is only
partial. A suitable energy cut that gives the same S/N ratio can
still be found, while some differences in the multipoles are now
visible, although well within typically expected errors. However,
both the study of single sources and of the energy spectrum of the
CGB should pin down the remaining uncertainty on $\tau$, and the
corresponding degeneracy should eventually be broken. GLAST, in
particular, will put strong constrains on the intensity of CIB
from the study of $\cal{O}$(10000) blazars expected to be detected
\cite{Chen:2004iz}.

Finally, let's note that these estimates are somewhat conservative:
summing the power at different $l$'s may favor the detection (see
e.g. \cite{Ando:2006mt}), and cross-correlating directly with the
maps we have produced would eventually rely on the whole
information. Indeed, to some extent also the information at small
scales can be exploited. In particular Fig. \ref{PSCzMaps} shows
that the anisotropy expected from small regions of the Virgo cluster
and along the Super-Galactic plane is significantly above the
average flux. Therefore, once the data on the CGB become available,
reassessing the problem with a detailed study will be mandatory.

\section{Summary and Conclusions}\label{conclusions}
The universe is pervaded by diffuse backgrounds of low-energy
photons, of cosmological origin like the CMB or due to stellar
activity, like the optical and infrared background (CIB). This
extragalactic background light makes the universe opaque to
energetic $\gamma$'s. The most energetic part of the gamma-ray
background (CGB) is thus of primary importance for high energy
astroparticle physics, since it acts as a ``cosmic calorimeter".
Besides telling us about the integrated history of the most powerful
astrophysical accelerators of the universe, it may constrain
non-standard physics taking place at high energy, even much higher
than the GeV$-$TeV scale. As an example, we remind that the most
stringent limits on the decays of superheavy particles coupled at
the tree-level {\it only to neutrinos} come from the observed
diffuse extragalactic $\gamma$-ray flux \cite{Berezinsky:2002hq}.

In this work we have studied the anisotropy  pattern of the CGB in
particular in the very-high energy regime, beyond about 100 GeV.
Due to the onset of the pair-production losses of VHE photons on
the CIB, most of the flux is coming from local structures, within
$z\alt 0.1$. Especially at the largest angular scales, the pattern
and {\it the amplitude} of the anisotropies are almost independent
of the source energy spectral shape and of the cosmological model:
modulo the magnitude of the CIB, the key parameter in shaping the
signature is the degree of the correlation of the gamma-emitters
with the known matter density field in the nearby universe (i.e.
their bias). For example, unless substructure emission dominates,
dark matter annihilation models for the origin of most of the CGB
predict a strong (quadratic) correlation of the flux with the
matter density, which should clearly manifest in the forthcoming
observations. It is interesting to note the nice complementarity
of observations made in the few GeV range with the ones in  the
VHE regime: for example around the GeV the statistics is large and
there is no uncertainty introduced by the absorption onto the CIB;
on the other hand, the VHE window is promising since, despite the
limited statistics, the anisotropies are larger and the
predictions are less dependent e.g. on the redshift evolution of
the sources. Moreover, the "filtering" of the emission of far
sources allows one to use more reliably the deterministic
information given by large scale structure catalogues, going
beyond merely statistical observables.

Starting from the PSCz astronomical catalogue, we produced maps of
the VHE gamma sky and estimated the potential of the satellite
mission GLAST or of extensive air shower observatories to detect
these features. The GLAST mission should be able to detect a
significant correlation of the diffuse gamma-ray emission with the
forecast maps presented in this work, providing an important
complementary observable to constrain the emission models. This is
especially true if an exotic contribution from dark matter
annihilation is relevant. Of course, before claiming an unambiguous
detection of dark matter in the CGB, detailed particle physics
models and a proper foreground analysis are required. Indeed, to
some extent strongly biased sources may mimick dark matter-like
features. Some early investigation of this issue are however quite
promising \cite{Ando:2006cr}.

EAS experiments are instead limited by the scarce cosmic ray
rejection capabilities, and only the next generation of instruments
like HAWC (or maybe already ARGO) may have real chances to achieve
the needed sensitivity. For EAS experiments, there is a further
remark. The Super-Kamiokande experiment has recently detected an
anisotropy at the level of few$\times 10^{-4}$ in the cosmic rays
around 10 TeV, from a sample of about $2\times 10^{8}$ muons
\cite{Guillian:2005wp} (MILAGRO has also detected this effect, as
mentioned in \cite{Atkins:2005wu}). At similar or higher energies,
the TIBET collaboration has reported the detection of several
anisotropies at the $\sim 0.1\%$ level in the cosmic ray flux,
probably associated with galactic sources and/or galactic transport
\cite{Amenomori:2006bx}. While the exposure needed to reveal the
features we have discussed so far is within the reach of the next
generation of EAS instruments, even assuming an excellent control
over experimental spurious effects, the ultimate limitation in
detecting these signatures comes from the understanding of the
intrinsic anisotropy in the CR background. Therefore, an efficient
gamma/hadron separation is not only necessary to enhance the
statistical significance of the point-like or diffuse gamma ray
sources observed by EAS instruments, but also to control
systematics. In particular, reversing the gamma cut and thus
enriching the sample in hadronic showers may help identifying and
removing non-gamma anisotropies.

\section*{Acknowledgments}
We thank the Danish Centre of Scientific Computing (DCSC) for
granting the computer resources used.  This work was also
supported by the PRIN04 "Fisica Astroparticellare" of Italian
MIUR. PS acknowledges support by the US Department of Energy and
by NASA grant NAG5-10842. HT was supported in part by NASA Grant
No. NNG05GG44G. SH and PS thank the Max-Planck-Institut f\"ur
Physik in Munich for hospitality and support during the initial
phase of this work. TH thanks the DARK Cosmology Centre for
hospitality during the course of this work. PS acknowledges the
members of the MAGIC group in Munich and of the GLAST group at
KIPAC/Stanford, and in particular Daniel Mazin and David Paneque,
for useful discussions. HT thanks Manoj Kaplinghat.

\appendix
\section{Gamma Propagation}\label{prop}
Here we report some details on the models used for the extragalactic
background light (EBL), and on the technique to account for
absorption effects in the propagation of photons. The main component
of the EBL is the cosmic microwave background, the spectral number
density of which is well known to obey a black-body spectrum

\begin{equation}
n_{\rm CMB}(\epsilon)=\frac{1}{\pi^2\,(\hbar
c)^3}\frac{\epsilon^2}{\exp \left( \epsilon/ T^0_{\rm CMB}
\right)-1}
\end{equation}
where $T^0_{\rm CMB}=2.73 {}\,$K=2.35${\times} 10^{-4}$eV is its
present temperature. For the Infrared/Optical Background (CIB),
which is the main source of gamma absorption, we use the simple
parametrization
\begin{equation}
n_{\rm CIB}(\epsilon)= \left\{
\begin{array}{l}
5.42{\times} 10^{11}{\rm eV}^{-1}{\rm cm}^{-3}\frac{\epsilon^{3.4}}{\exp(\epsilon/T_F)-1} \:\:\:\:\:\:\:\:\:\:\:\:\lambda\in 200\div 2000\,\mu{\rm m}, \\
7.4{\times} 10^{-4}{\rm eV}^{-1}{\rm cm}^{-3}\epsilon^{-2.295}\:\:\:\:\:\:\:\:\:\:\:\:\:\:\:\:\:\:\:\:\:\:\lambda\in 6.0\div 200\,\mu{\rm m}, \\
7{\times} 10^{-3}{\rm eV}^{-1}{\rm
cm}^{-3}\frac{e^{-\epsilon}}{\epsilon}\:\:\:\:\:\:\:\:\:\:\:\:\:\:\:\:\:\:\:\:\:\:\:\:\:\:\:\:\:\:\:\:\lambda<6.0\,\mu{\rm
m},
\end{array}\right.
\label{CIBpar}\end{equation}
where $T_F=13.6\,$K; and the $\lambda-\epsilon$ conversion factor
is given by $\lambda(\mu{\rm m})=1.24/\epsilon$(eV). This is a
conservative estimate of the CIB consistent with data and
constraints reported in \cite{Hauser:2001xs}.

\begin{figure}[!tb]
\begin{center}
\begin{tabular}{c}
\epsfig{file=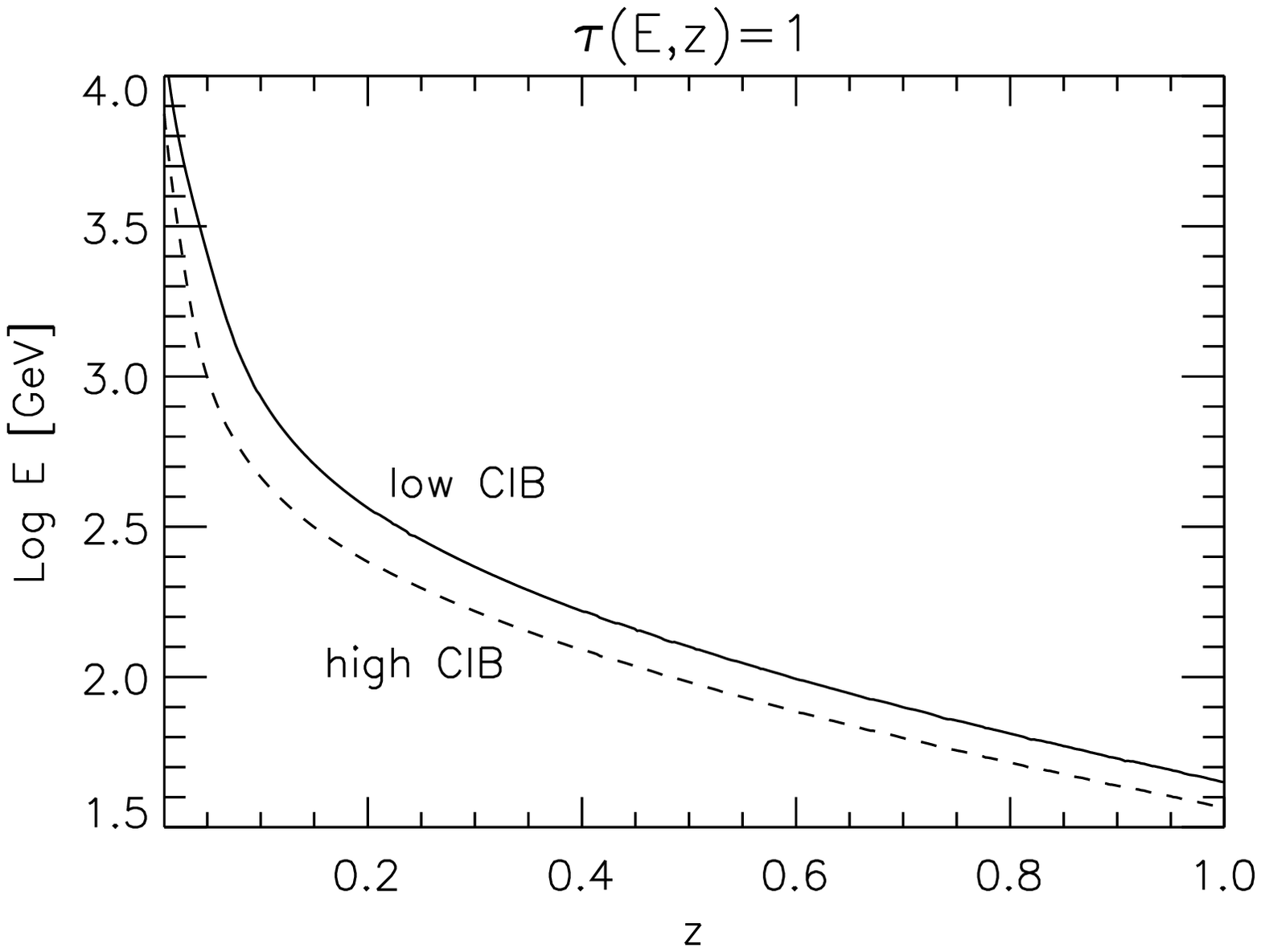,width=12cm}
\end{tabular}
\end{center}
\caption{Critical optical depth $\tau=1$ in function of $\gamma$-ray
energy and redshift. The low CIB case (dashed line) corresponds to
Eq. (\ref{CIBpar}), while the high CIB case corresponds to twice
that value.}\label{taucurve}
\end{figure}

The time evolution of the backgrounds is obtained simply by
redshifting, i.e.
\begin{equation}
  n(\epsilon,z)= (1+z)^2 n_0 \left[ \frac{\epsilon}{1+z} \right].
\end{equation}
This result is exact for the CMB that is a truly primordial
background, while for CIB one should in principle perform a
simulation of the star formation and dust clustering that produced
this background in the recent past. However accurate modelling of
this process suggests that most of the background formed in a burst
at $z\simeq 4-5$ near the peak of star formation rate, so that for
our purposes (propagation till $z\simeq 0.2$) simple redshifting is
quite accurate \cite{Hauser:2001xs}.

In principle, the interaction of high energy photons with the EBL
is a complex process in which the $e^-e^+$ created through pair
production interact again with background photons via inverse
Compton scattering producing new high energy $\gamma$'s. We have
developed our own code for calculating the evolution of gamma
cascades, along the lines of Ref.~\cite{Lee:1996fp} (which we
refer to for the details, including the general formalism of
cascade propagation). Luckily, however, the complicated regime
where the cascade of high energy $e^+,e^-,$ and $\gamma$ has to be
followed in details for calculating the final energy distribution
is only attained for energies $>10^{15}\,$eV. In the TeV range
relevant for gamma astronomy the development of the shower via
secondary particle dynamics can be safely neglected, and the
observed flux can be calculated by simply considering an
energy-dependent depletion factor for the spectrum in addition to
the effect of redshifting. Thus, if the spectrum at the source is
of the form $g(E)$, the observed spectral shape of the signal will
be
\begin{equation}
g(E_\gamma)\propto
g[E_\gamma(1+z)]\mathcal{P}(E_\gamma,z),\label{spectrabs}
\end{equation}
where $E_\gamma$ is the energy we observe today and
$\mathcal{P}(E_\gamma,z)$ is the probability for a photon emitted at
redshift $z$ to survive without interacting till now, when it
reaches us with energy $E_\gamma$. This probability is written as
\begin{equation}
\mathcal{P}(E_\gamma,z)\equiv e^{-\tau(E_\gamma,z)},
\end{equation}
where the optical depth $\tau$ is
\begin{equation}
\tau(E_\gamma,z)\equiv \int_0^z\d z' \frac{c}{(1+z')H(z')} \int
\d\epsilon\,n(\epsilon,z') \int \d\mu \frac{1- \mu}{2} \sigma_{\rm
PP} (E_{\gamma}(1+z'),\epsilon,\mu),
\end{equation}
that is the rate of pair production of photons on the EBL
integrated over the time during propagating from redshift $z$ to
0. The function $\sigma_{\rm PP} (E,\epsilon,\mu)$ is the
theoretically and experimentally well-known pair production cross
section of a photon of energy $E$ impinging over a background
photon of energy $\epsilon$ with a cosinus of the impact angle
given by $\mu$. In Fig.~\ref{taucurve} we show the derived
critical $\tau(E_\gamma,z)=1$ contour that represents the gamma
redshift horizon as a function of the energy. Note that the EBL is
known only within a factor of O(2). As we show in
Fig.~\ref{taucurve} with the case of a doubled optical depth,
there is a degeneracy between the energy horizon and the true
value of $\tau$. This translates into a degeneracy between the
anisotropy pattern and the assumed intensity of the CIB. However,
both the study of single sources and of the energy spectrum of the
CGB should pin down the remaining uncertainty on this quantity,
and the corresponding degeneracy should eventually be broken.
Given existing uncertainties, we find a satisfactory agreement of
the function $\tau$ we computed with more detailed studies
including the CIB distribution and evolution from simulations of
star and galaxy formation~\cite{Stecker:2005qs}.

\section{Summary of the properties of noise}\label{ShotNoise}
In this Appendix we briefly review the noise properties of a
discrete poisson process on the sphere. The main results can be
easily obtained analytically and concern the amplitude of the noise
variance and the related power spectrum. The map resulting from a
random realization of $N$ equally weighted points can be written as
$f(\hat{\Omega})=4 \pi /N \sum_i
\delta(\hat{\Omega}-\hat{\Omega}_i)$, normalized to its mean value,
so that to be adimensional; then the harmonic expansion coefficients
$a_{lm}=\int d\hat{\Omega}f Y_{lm}( \hat{\Omega})$ follow (apart the
constant monopole contribution) a gaussian distribution with
$\langle a_{lm}\rangle =0$ and
\begin{equation}\label{almsigma}
    \sigma_{a_{lm}}^2=\langle a_{lm}^2\rangle = \frac{4\pi}{N}
\end{equation}
\emph{independent} from $l,m$. From this, the spectrum of the shot
noise $C_l=\sum_m |a_{lm}|^2/(2 l+1)$ follows a $\chi^2$
distribution with mean
\begin{equation}\label{clmean}
    \langle C_l\rangle= \frac{4\pi }{N}
\end{equation}
again independent of $l$, and variance
\begin{equation}\label{clsigma}
    \sigma_{C_l}^2 = \langle C_l^2\rangle- \langle C_l\rangle^2 =\left(\frac{4 \pi}{N}\right)^2  \frac{2}{2l+1}.
\end{equation}

These results surely hold in the limit of large statistics;
however for our applications it is worth testing them also in the
limit of very low statistics, say for $N\le 1000$. We performed a
set of Monte Carlo simulations to clarify the issue an found that
the simulations and analytic results are in perfect agreement even
in the completely unphysical limit of $N=10$; the analytical
results can then be safely used in all the cases of interest.

In general, in addition to the white noise $C_l^N$, one has a
signal $C_l^S$. Moreover, there is normally incomplete sky
coverage, and additional white noise may be present, as in our
case because of the background due to cosmic rays passing the
cuts. A generalization of Eq. (\ref{clsigma}) then reads
\begin{equation}
\sigma_{C_l}^2=\frac{2}{(2l+1)f_{\rm sky}}(C_l^S+ C_l^N)^2,
\end{equation}
where $f_{\rm sky}$ is the fraction of the sky accessible to the
experiment (assumed with uniform acceptance over this region), and
the noise spectrum $C_l^N$ (including the cosmic ray background) is
given by
\begin{equation}\label{shotnoiselevel}
C_l^N=\frac{4\pi f_{\rm sky}}{N_\gamma}\left[1+\frac{N_{\rm
CR}}{N_{\gamma}}\right].
\end{equation}

For the $a_{lm}$'s, the variance due to the shot-noise plus the
background is written as
\begin{equation}\label{sigmashotnoiselevel}
  \sigma_{a_{lm}}^2=C_l^N=\frac{4\pi f_{\rm sky}}{N_\gamma}\left(1+\frac{N_{\rm
CR}}{N_{\gamma}}\right).
\end{equation}
In principle, the complete formulae receive a correction at large
$l$ due to the finite angular resolution of the experiment that is
easily implemented performing everywhere the substitution $C_l^N
\longrightarrow C_l^N \exp(l^2 \sigma_b^2/2)$ $\sigma_b$ being the
angular resolution of the experiment. However, since the
experimental resolutions are better than a degree, and we limit
our considerations to the most prominent signatures at large
scales, this correction is unnecessary for our application.

We have used these results in estimating the errors reported in Sec.~\ref{detection}.
Note that what is really measured is always the sum
signal+noise: the noise is an unavoidable component in any
experiment. However, being constant in $l,m$, the average level of
the noise can be fitted and subtracted (this is trivial for
deterministic predictions since $\langle a_{lm}^N\rangle=0$). On the
other hand, the error in its determination depends on the
sensitivity of the experiment and thus the statistics collected, and
on the level of background rejection.

\section*{References}

\end{document}